\documentclass[aps,prb,showpacs,twocolumn,reprint,amsmath,amssymb]{revtex4-1}
\usepackage{amsmath}
\usepackage{graphicx}
\usepackage{color}
\usepackage{multirow}
\usepackage{dcolumn}
\usepackage{textcomp}
\begin{document}
\title{Effects of interface oxygen vacancies on electronic bands of
FeSe/SrTiO$_3$(001)}
\author {M. X. Chen}
\affiliation{Department of Physics, University of Wisconsin--Milwaukee,
Milwaukee, WI 53201}
\author {Zhuozhi Ge}
\affiliation{Department of Physics, University of Wisconsin--Milwaukee,
Milwaukee, WI 53201}
\author {Y. Y. Li}
\affiliation{Department of Physics, University of Wisconsin--Milwaukee,
Milwaukee, WI 53201}
\author {D. F. Agterberg}
\affiliation{Department of Physics, University of Wisconsin--Milwaukee,
Milwaukee, WI 53201}
\author {L. Li}
\affiliation{Department of Physics, University of Wisconsin--Milwaukee,
Milwaukee, WI 53201}
\author {M. Weinert}
\affiliation{Department of Physics, University of Wisconsin--Milwaukee,
Milwaukee, WI 53201}

\date{\today}

\begin{abstract}
Modifications of the electronic bands of thin FeSe films due to oxygen
vacancies in the supporting SrTiO$_3$(001) substrate -- and the
interplay with spin-orbit coupling, magnetism, and epitaxy
-- are investigated by first-principles supercell
calculations.  Unfolded ($k$-projected) bands show that the oxygen
vacancies both provide electron doping to the interface FeSe layer and
also have notable effects on the details of the bands around the Fermi
level, including renormalizing the width of the Fe-3$d$ band near the
Fermi level by a factor of about 0.6, and causing a splitting of
$\sim$40 meV at the M point for the checkerboard antiferromagnetic
configuration. For an FeSe bilayer, the modifications to the bands are
mainly limited to the interface FeSe layer.  While spin-orbit-coupling
induced band splittings of $\sim$30 meV at M for the ideal
FeSe/SrTiO$_3$(001) interfaces are comparable to the splitting due to
oxygen vacancies, the effects are not simply additive. Calculations and
comparison to our scanning tunneling microscopy images of MBE-grown FeSe films on
SrTiO$_3$(001) suggest that a common
defect may be Se bound to an oxygen vacancy at the interface, 
\end{abstract}


\maketitle
\section{Introduction}
Recently, the discovery of high T$_c$ superconductivity in FeSe
monolayers grown on SrTiO$_3$(001) (STO) has inspired intense interest
in the FeSe/STO interface and FeSe thin
films.\cite{qing-yan_2012,huang_2015,zhou_2015,zhang_Ding_2015,miyata_2015,shiogai_2016}
By means of in situ scanning tunneling microscopy (STM) Wang \textit{et
al.} observed a superconducting gap of about 20 meV for epitaxially
grown FeSe monolayers after proper annealing, suggesting that the
superconducting T$_c$ in the grown FeSe monolayers may be as high as 77
K.\cite{qing-yan_2012} Subsequent transport measurements and
angle-resolved photoemission spectroscopy (ARPES) experiments also
observed high-T$_c$ superconductivity in this
system.\cite{zhang_direct_2014,sun_high_2014,ge_2015,liu_electronic_2012,he_phase_2013,tan_2013,liu_dichotomy_2014}
One puzzling feature of these observations is that the Fermi surface of
the superconducting monolayer FeSe/SrTiO$_3$ is characterized by an
electron-like pocket centered around the M points, while no pocket
appears near the $\Gamma$
point.\cite{liu_electronic_2012,he_phase_2013,tan_2013,liu_dichotomy_2014}
This challenges the Fermi surface nesting scenario that relies on
nesting between M and $\Gamma$-point electronic states.
\cite{mazin_2008,kuroki_2008}

Although the mechanism of the high T$_c$ superconductivity in the FeSe
monolayers is under debate, there is consensus that the substrate plays
an important role. For example, the induced strain on the FeSe films
modifies the electronic properties.\cite{peng_tuning_2014} In addition,
it has been argued that the coupling of ferroelectric phonons in the
substrate to electrons in FeSe may enhance the
superconductivity.\cite{lee_interfacial_2014} Most importantly, the
electronic properties of the grown FeSe monolayers are strongly
dependent on annealing, i.e., the semiconducting as-grown FeSe monolayer
becomes metallic upon proper annealing and then superconducting with
further annealing.\cite{zhang_Xue_2014} However, FeSe multilayers show
distinct differences in that they remain non-superconducting even after
the annealing.  This suggests that oxygen vacancies formed during
annealing play an important role in the evolution of the electronic
properties.

There have been a number of first-principles studies of both
free-standing and supported FeSe thin films,
\cite{bazhirov_2013,liu_atomic_2012,bang_atomic_2013,cao_interfacial_2014,xie_oxygen_2015,
shanavas_2015,zheng_band_2015,berlijn_2014,zheng_2013,liu_2015}
investigating the effect of the SrTiO$_3$ substrate,
\cite{liu_atomic_2012,bang_atomic_2013,cao_interfacial_2014,xie_oxygen_2015,shanavas_2015}
surface adatoms,\cite{shanavas_2015,zheng_band_2015} Se-vacancies,
\cite{berlijn_2014} electron doping,\cite{zheng_2013,liu_2015} and
oxygen vacancies
(O-vac).\cite{bang_atomic_2013,cao_interfacial_2014,xie_oxygen_2015,shanavas_2015}
It is found that there is no strong chemical bonding between the FeSe
monolayer and perfect SrTiO$_3$(001), although the Se atoms in the FeSe
monolayer do prefer the top sites of the cation atoms of the substrate.
\cite{liu_atomic_2012} The binding is enhanced by interface oxygen
vacancies, which dope electrons into the FeSe monolayer, consequently
modifying the Fermi surface of the
overlayer.\cite{bang_atomic_2013,cao_interfacial_2014,xie_oxygen_2015}
The effect of oxygen vacancies on the bands of the nonmagnetic state has
been investigated within the virtual crystal approximation, showing that
the hole pockets around $\Gamma$ are not fully
removed.\cite{shanavas_2015} However, thus far, the  understanding of
the effects of O vacancy at the interface between FeSe thin films and
the SrTiO$3$(001) substrate remains incomplete, including issues related
to the effect of oxygen vacancies on the electronic structures of FeSe
layers both at the interface and farther away.

\begin{figure}
  \includegraphics[width=0.38\textwidth]{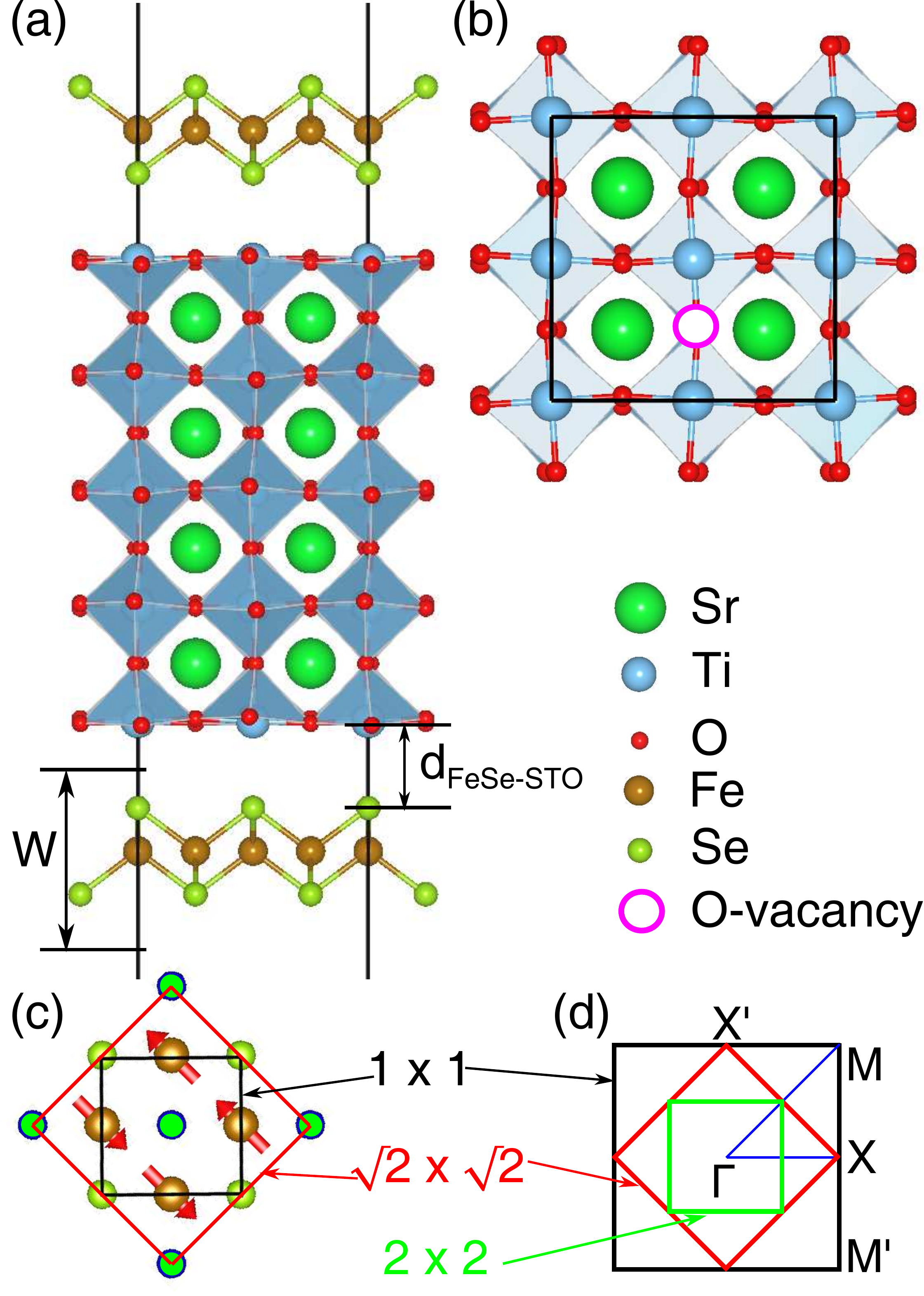}
  \caption{(color online) Structural model of FeSe/SrTiO$_3$(001).
  (a) Side view of monolayer FeSe/SrTiO$_3$(001) in a 2$\times$2
supercell.  d$_\textrm{FeSe-STO}$ is the planar distance between the
interface Ti and Se atoms.
  (b) Top view of 2$\times$2 SrTiO$_3$(001) with one O-vac.  W
represents the spatial window in which supercell wave functions are
projected onto the (1$\times$1) FeSe crystallographic (two Fe) cell.
  (c) Magnetic unit cell for CL-AFM denoted by the red box, which is in
a $\sqrt{2}\times\sqrt{2}$ supercell.  The crystallographic unit cell is
represented by the black box.
  (d) Brillouin zones (BZs) of FeSe in the 1$\times$1 unit cell,
$\sqrt{2}\times\sqrt{2}$ and 2$\times$2 supercells.
  }
 \label{fig1}
\end{figure}

In this paper, we investigate the effects of interface O-vac on the
electronic bands of FeSe monolayers and bilayers supported on
SrTiO$_3$(001) by carrying out first-principles calculations.  The
O-vacant systems are modeled by supercells, and the unfolded bands are
then obtained by projecting the supercell wave functions onto the
1$\times$1 unit cell.  It is found that the interfacial oxygen vacancy
not only provides electron doping to the interface FeSe layer, but also
has notable effects on the band details.  In particular, for the
checkerboard (CB) antiferromagnetic (AFM) state, the oxygen vacancies
induce a band splitting at the M point (and along X-M), and lowers the
Fe-3$d_{z^2}$ states at the $\Gamma$ point, thus significantly
renormalizing the width of the Fe-3$d$ band near the Fermi level.
However, the effects of the O-vacancy on the electronic properties of
the top layer of bilayer FeSe are limited.  In addition, spin-orbit
coupling (SOC) can induce splittings at M comparable to those due to
oxygen vacancies, demonstrating the possible importance of SOC in
understanding this class of materials.\cite{soc-2015}

\section{Structural Model and Computational Details}

As shown in Fig.~\ref{fig1}, the SrTiO$_3$(001) substrate (with a
lattice constant of 3.905 \AA) was modeled by a TiO$_2$-terminated slab
consists of five TiO$_2$ layers and four SrO layers, based on the
relaxed structure of the antiferrodistortive SrTiO$_3$ bulk.  To avoid
dipole interactions between slabs, a single layer of FeSe is
symmetrically placed on each side of SrTiO$_3$(001).  Two epitaxial
relationships for the FeSe/SrTiO$_3$(001) interface were considered: The
bottom Se atoms sit directly above ($i$) the surface Ti atoms (Type A),
which was previously found to be more favorable compared to other
configurations;\cite{liu_atomic_2012} or ($ii$) the surface O atoms
(Type B). Although the ideal Type B interface is not energetically
favorable, if there is a strong enough interaction between the oxygen
vacancy and the Se, it is at least plausible that this epitaxial
relationship could be nucleated; this issue is addressed below.  The
various FeSe/SrTiO$_3$(001) slabs are separated from their periodic
images by $\sim$20 \AA{} vacuum regions. An oxygen vacancy is modeled by
removing one surface oxygen atom in a 2$\times$2 (or
$\sqrt{2}$$\times$$\sqrt{2}$) supercell (STO-vac; see
Fig.~\ref{fig1}(b)).  The calculations were performed using the Vienna
Ab initio Simulation Package.
\cite{kresse_efficiency_1996,kresse_efficient_1996} The exchange
correlation functional is approximated by the generalized gradient
approximation as parametrized by Perdew, Burke and
Ernzerhof,\cite{perdew_1996} and the pseudopotentials were constructed
by the projector augmented wave
method.\cite{bloechl_projector_1994,kresse_ultrasoft_1999} van der Waals
(vdW) dispersion forces between the adsorbate and the substrate were
included using the vdW-DF method developed by Klime\v{s} and
Michaelides.\cite{PhysRevB.83.195131} An 8$\times$8 Monkhorst-Pack
$k$-mesh was used to sample the surface BZ and a plane-wave energy cut
off of 400 eV was used for structural relaxation and electronic
structure calculations.  For the structural relaxation only the FeSe and
the top TiO$_2$ and SrO layers were allowed to relax, with a threshold
of 0.001 eV/\AA{} for the residual force on each atom while other atoms
were held fixed.  Three different magnetic configurations of the FeSe
films were considered: nonmagnetic (NM), checkerboard antiferromagnetic
(CB-AFM), and collinear antiferromagnetic (CL-AFM); the unstable
ferromagnetic state is not considered. As shown in Table~\ref{table1}, the CL-AFM
configuration is calculated to be lower in energy than the CB-AFM, with the
difference decreasing in the presence of oxygen vacancies.  Although the FeSe may not have
long-range magnetic order, these ordered calculations do provide insight
into the effect of short-range magnetic order, which can have
significant effects on the bands, i.e., non-magnetic calculations are
not good representations of paramagnetic systems if there are local
moments. For the Fe-chalcogenides materials, even in the paramagnetic
state, the NM calculated bands have Fermi surfaces and dispersions in
disagreement with experiments;\cite{soc-2015} we include the NM results
here for comparison purposes even though these are not expected to
correspond to the physical system.

\begin{table}[b]
\begin{center}
\begin{tabular}[c]{cccccccc}
\hline
\hline
  & \multicolumn{3}{c}{ideal} &  \quad &  \multicolumn{3}{c}{STO-vac} \\
                &  NM  & CB-AFM & CL-AFM  &   &  NM  & CB-AFM
& CL-AFM    \\
\cline{2-4}                        \cline{6-8}
  E$_{tot}$            & 0.00 &-0.40  &-0.55    &   & 0.00 &-0.43      &-0.51      \\
  d$_{\rm FeSe-STO}$ & 3.00 & 2.98  & 3.00    &   & 2.82 & 2.80      & 2.82      \\
  d$_{\rm Fe-Se}$    & 2.34 & 2.40  & 2.41    &   & 2.34 & 2.40      & 2.41      \\
  $\mu$ ($\mu_B$)   &      & 2.11   & 2.40    &   &      & 2.04--2.15 &
2.34--2.45      \\
\hline
\hline
\end{tabular}
\end{center}
\caption{(color online) Structural properties of the FeSe/SrTiO$_3$(001)
Type A interfaces, both ideal and with an O vacancy, for different
magnetic configurations.  
E$_{tot}$ (in eV per FeSe unit cell) is the energy difference between magnetic states and the nonmagnetic state. 
d$_{\rm FeSe-STO}$ (see Fig.~\ref{fig1}) and
d$_{\rm Fe-Se}$ (in \AA) are the interlayer distance and Fe-Se bond
length, respectively; $\mu$  is the magnitude of the Fe magnetic
moments.}
\label{table1}
\end{table}

\begin{figure*}
  \includegraphics[width=0.90\textwidth]{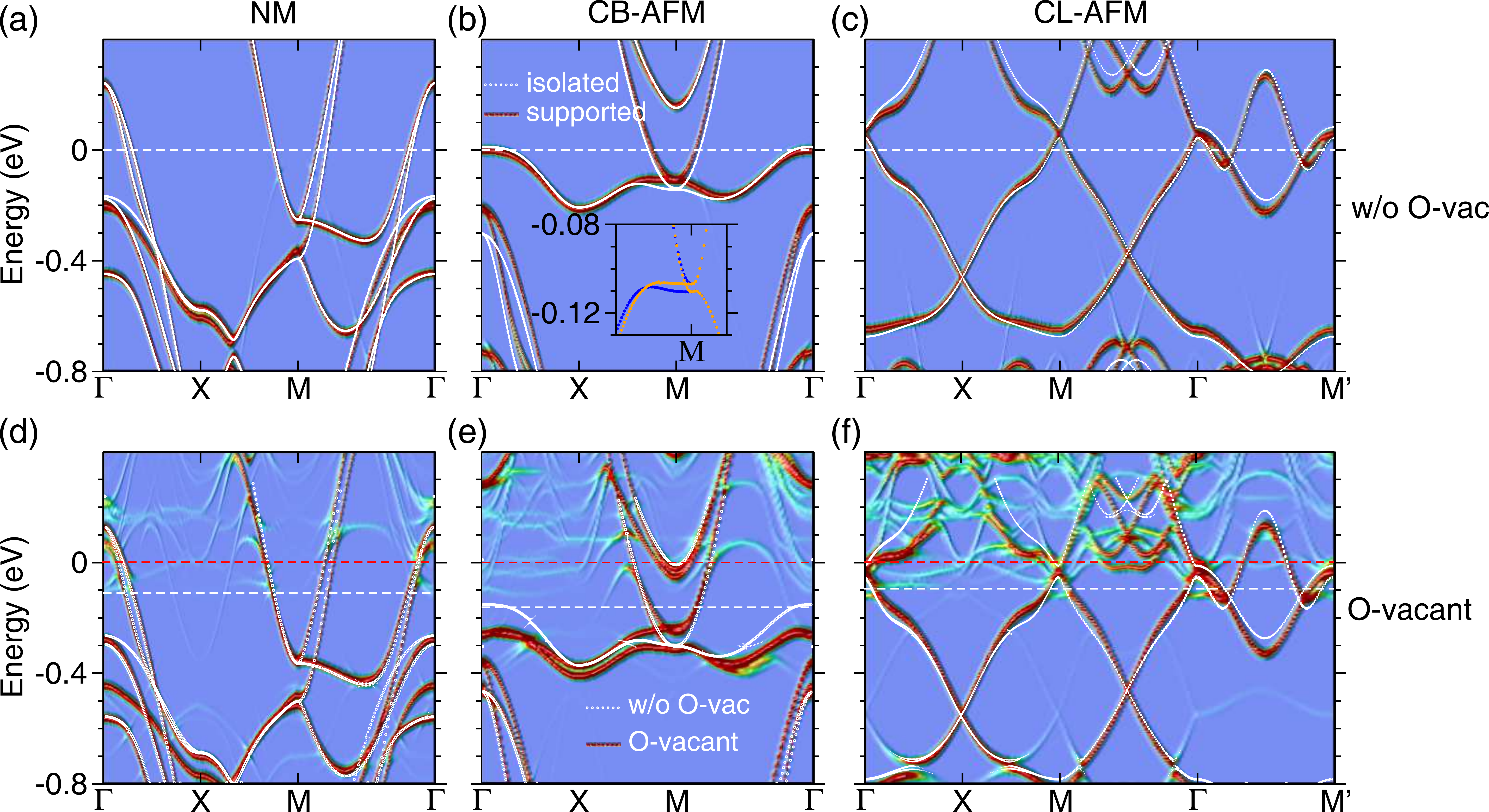}
  \caption{(color online) $k$-projected bands for FeSe monolayers on
SrTiO$_3$(001).  Ideal interface for (a) NM, (b) CB-AFM (inset: expanded
scale spin-resolved bands around M), and (c) CL-AFM magnetic
configurations.
  (d)-(f) Corresponding bands for oxygen vacancy.  Bands for isolated
FeSe layers with the same internal (relaxed) structure (white) are
overlaid for comparison.  White and red dashed lines denote the Fermi
levels of the isolated and supported FeSe, respectively.
  }
 \label{fig2}
\end{figure*}

The calculated electronic bands of the supported FeSe thin films used
supercells, leading to band folding.  To discern the effects of the
interface O-vacancy on the electronic bands of the FeSe overlayer (or
different magnetic orderings), the supercell wave functions in FeSe
(spatial window W shown in Fig.~\ref{fig1}) were projected onto the
corresponding $k$ of the (1$\times$1) FeSe cell using the layer
$k$-projection technique.\cite{bufferlayer,Goodwin_39,kproj_88,chen_revealing_2014}
This technique also allows for a more direct comparison to the
photoemission results.

\section{Results and Discussion}

\subsection{Se-Ti (Type A) interface}

Table~\ref{table1} summarizes the calculated structural properties of
the interfaces of an FeSe monolayer with the ideal SrTiO$_3$(001)
substrate and with STO-vac.  The interlayer distance d$_{\rm FeSe-STO}$
remains almost unchanged for all the considered magnetic states.  (The
surface oxygen atoms are $\sim$0.08 \AA{} above the Ti atoms.) The value
of 2.98 \AA{} for CB-AFM is slightly smaller than the 3.06 \AA{}
reported previously\cite{liu_atomic_2012,zheng_2013} as a result of the
vdW correction, while the 3.0 \AA{} for the CL-AFM state is consistent
with Shanavas and Singh's calculation.\cite{shanavas_2015} The Fe-Se
bond lengths for the NM state are slightly shorter than those for the
magnetic states.  Although the two Se atoms are inequivalent because of
the presence of the substrate, there is negligible difference in d$_{\rm
Fe-Se}$ for the two Se layers, implying that symmetry breaking in the
supported FeSe monolayer due to the substrate is rather weak.
Introduction of an O vacancy affects d$_{\rm FeSe-STO}$ significantly
reduces the interlayer separation d$_{\rm FeSe-STO}$ by about 10\%, in
agreement with a previous study.\cite{xie_oxygen_2015} However, the
effect on the internal structure of the FeSe overlayer is negligible:
d$_{\rm Fe-Se}$ remains almost unchanged and the Fe atoms are still
nearly co-planar.  The induced variations of the Fe magnetic moments are
in the range of 0.04-0.07 $\mu_B$.

\begin{figure}[b]
  \includegraphics[width=0.3\textwidth]{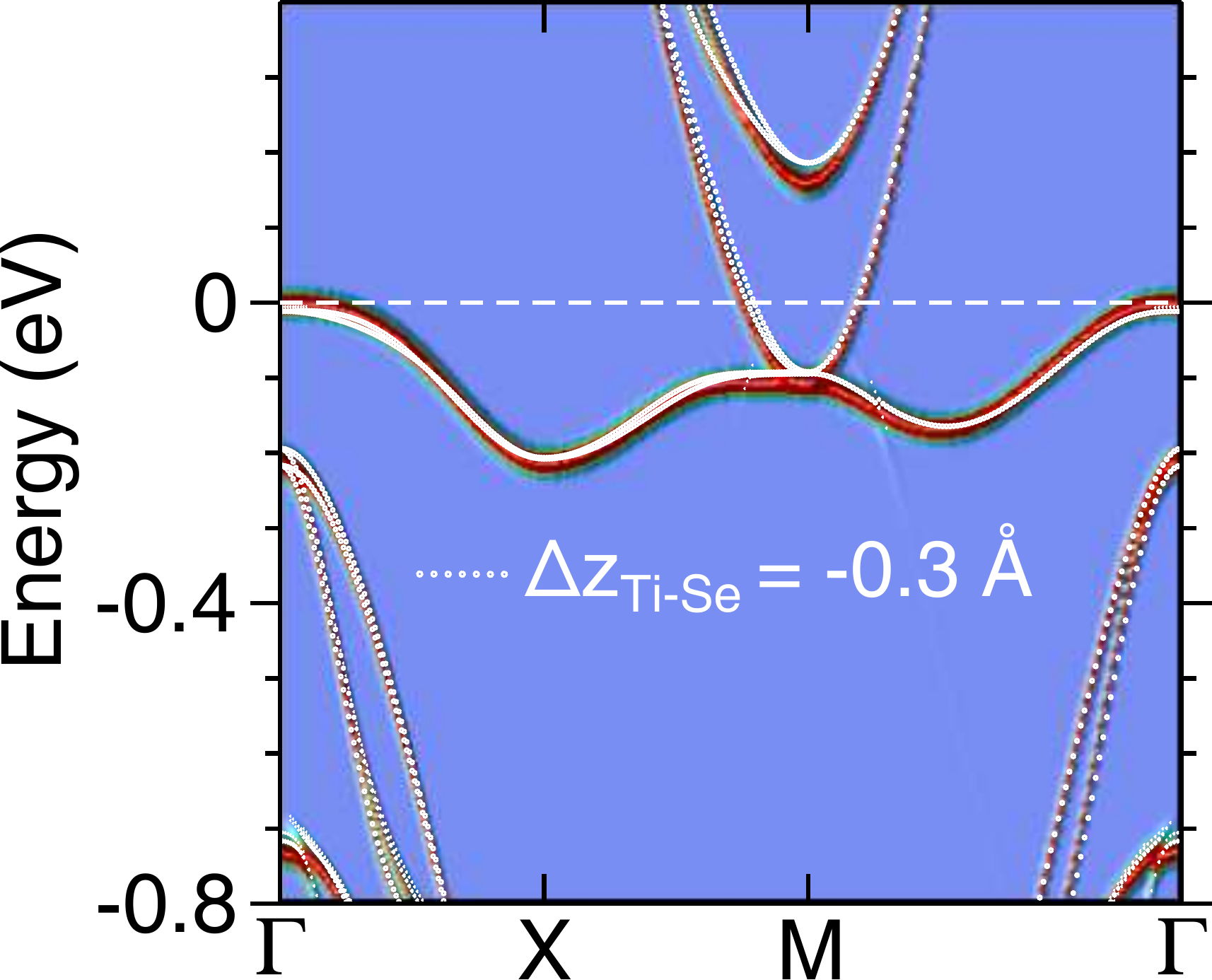}
  \caption{(color online) Comparison of bands for the perfect
FeSe/SrTiO$_3$(001) at the equilibrium interlayer distances and at a
reduced (by 0.3 \AA) interlayer separation.
  }
 \label{fig3}
\end{figure}

Figure ~\ref{fig2} shows $k$-projected bands for supported FeSe in the
different magnetic states, both with and without the interface
O-vacancy. Bands of the isolated FeSe monolayers with the same structure
as in the supported case (i.e., with the substrate removed) are overlaid
for comparison.  Without the interface oxygen vacancy,
Figs.~\ref{fig2}(a)-(c), the bands of the supported FeSe are similar to
those of the free-standing FeSe monolayer.  For the NM state, there are
two hole bands near $\Gamma$ and two electron bands around M crossing
the Fermi level (E$_F$).  For the CB-AFM state there is a hole band with
a top that is slightly higher than E$_F$ at $\Gamma$ and a electron band
around M, which is spin-degenerate.  The bands at $\Gamma$ near E$_F$
due to Fe-3$d$ orbitals are pushed below E$_F$ if a Hubbard-U correction
is added.\cite{zheng_2013} For the CL-AFM state, there are hole bands
crossing E$_F$ near both $\Gamma$ and M, in agreement with a previous
study.\cite{liu_atomic_2012} Figures~\ref{fig2}(a)-(c) also indicate
that the ideal substrate has minor effects on the bands of FeSe near the
Fermi level and a negligible shift of the Fermi level.  For the NM
state, the FeSe bands remain almost unaffected, except for a small shift
of the states about $\Gamma$ near $-$0.2 eV and the states about M near
$-$0.4 eV.  For the CB-AFM state, bands at $\Gamma$ near $-$0.3 eV are
shifted upward by about 0.1 eV.

\begin{figure}[t]
  \includegraphics[width=0.48\textwidth]{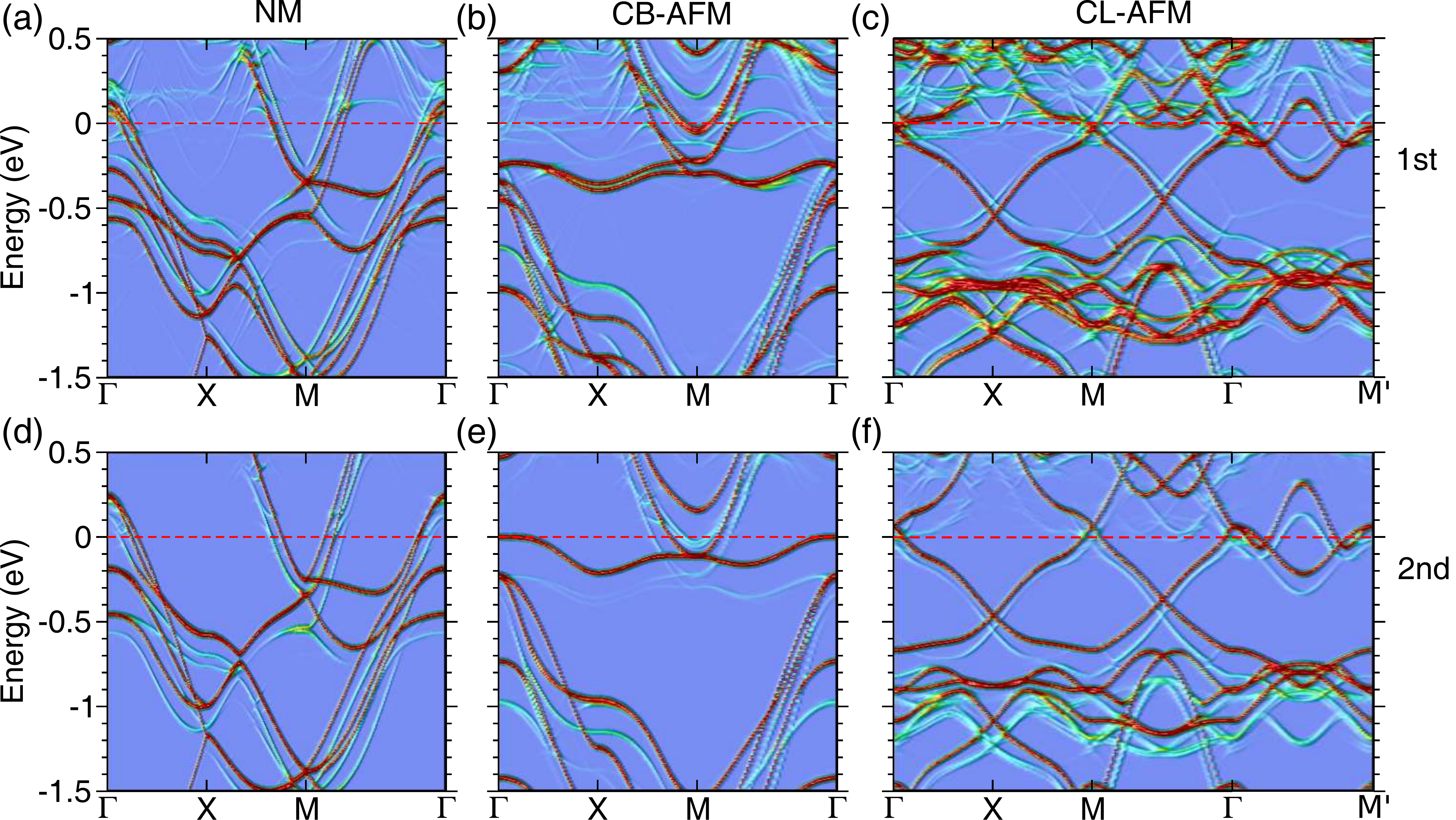}
  \caption{(color online) $k$-projected bands for bilayer FeSe on
SrTiO$_3$(001) with an O vacancy in different magnetic configurations
for (a)-(c) the interface and (d)-(f) top FeSe layer in the NM, CB-AFM,
and CL-AFM states, respectively.
  }
 \label{fig4}
\end{figure}

Figures~\ref{fig2}(d)-(f) indicate that there is a significant shift of
the Fermi level as a result of electron doping provided by the
O-vacancy.  Accordingly, the electron pockets near $\Gamma$ become
smaller and the hole pocket about M becomes larger for the NM state.
However, the main profile of FeSe bands is basically maintained, except
for the bands about $\Gamma$ at about $-$0.3 eV composed of
Fe-3$d_{xz,z^2}$, which are shifted downward to $-$0.45 eV.  For the
CB-AFM state the O-vacancy not only shifts E$_F$ but also lowers the
band at $\Gamma$.  As a result, the hole pocket at $\Gamma$ disappears
and the band-width is renormalized by a factor of about 0.6.  Moreover,
it lifts the spin degeneracy of bands in between -0.3 ev and -0.2 eV
along $\Gamma$-X-M, inducing a band splitting of about 40 meV at the M
point.  For the CL-AFM state, the pockets near $\Gamma$ and near M are
hole-like.  The empty states of FeSe are strongly perturbed by the
hybridization with the substrate.

The significant difference in the electronic bands for CB-AFM
configuration with and without the O-vacancy is related to the
structural differences.  Table \ref{table1} shows that the structure of
FeSe thin films remains almost unaffected, but the interlayer distance
is considerably reduced in the presence of the O-vacancy, which may
strengthen the interaction between the overlayer and the substrate,
including inducing more charge transfer between them.  To study the
effect of the interlayer separation, calculations were done for ideal
FeSe/SrTiO$_3$(001) with the interlayer distance reduced by 0.3 \AA.
Figure~\ref{fig3} indicates that the FeSe bands are basically
unaffected.  Therefore, the dramatic changes in the bands of FeSe can be
attributed to the O-vacancy, which both dopes electrons to the FeSe
overlayer and  pins the Fermi level of the FeSe to the substrate band
gap.  Indeed, previous calculations have demonstrated that bands at M
and $\Gamma$ for the freestanding CB-AFM FeSe are sensitive to charge
doping.\cite{zheng_2013}

The bands for an FeSe bilayer on SrTiO$_3$(001) with an oxygen vacancy
are shown in Fig.~\ref{fig4}.  One important feature that can be seen
from Fig.~\ref{fig4} is that the $k$-projected bands for the interface
layer are similar to those for monolayer FeSe/STO-vac, while the bands
for the top layer (Figs.~\ref{fig4}(d)-(f)) resemble those for the
perfect FeSe/STO.  In particular, a comparison of Figs.~\ref{fig4}(b)
and (e) indicates that for the CB-AFM state, the band splitting at M
point is almost negligible for the top layer and a pocket appears at
$\Gamma$.  For the CL-AFM state, the strong band hybridization between
empty states of the interface FeSe layer and the substrate along
M-$\Gamma$ (Fig.~\ref{fig4}(c)) is not seen for the top layer.  This
indicates that the O-vacancy has little effect on the top layer, which
stems from the interface FeSe layer effectively screening the charges
created by the interface O-vacancy.  This is consistent with previous
calculations that find that charge induced by the O-vacancy are mainly
distributed at the interface.\cite{cao_interfacial_2014}

\subsection{Dependence on oxygen-vacancy concentration}

\begin{figure} 
  \includegraphics[width=0.30\textwidth]{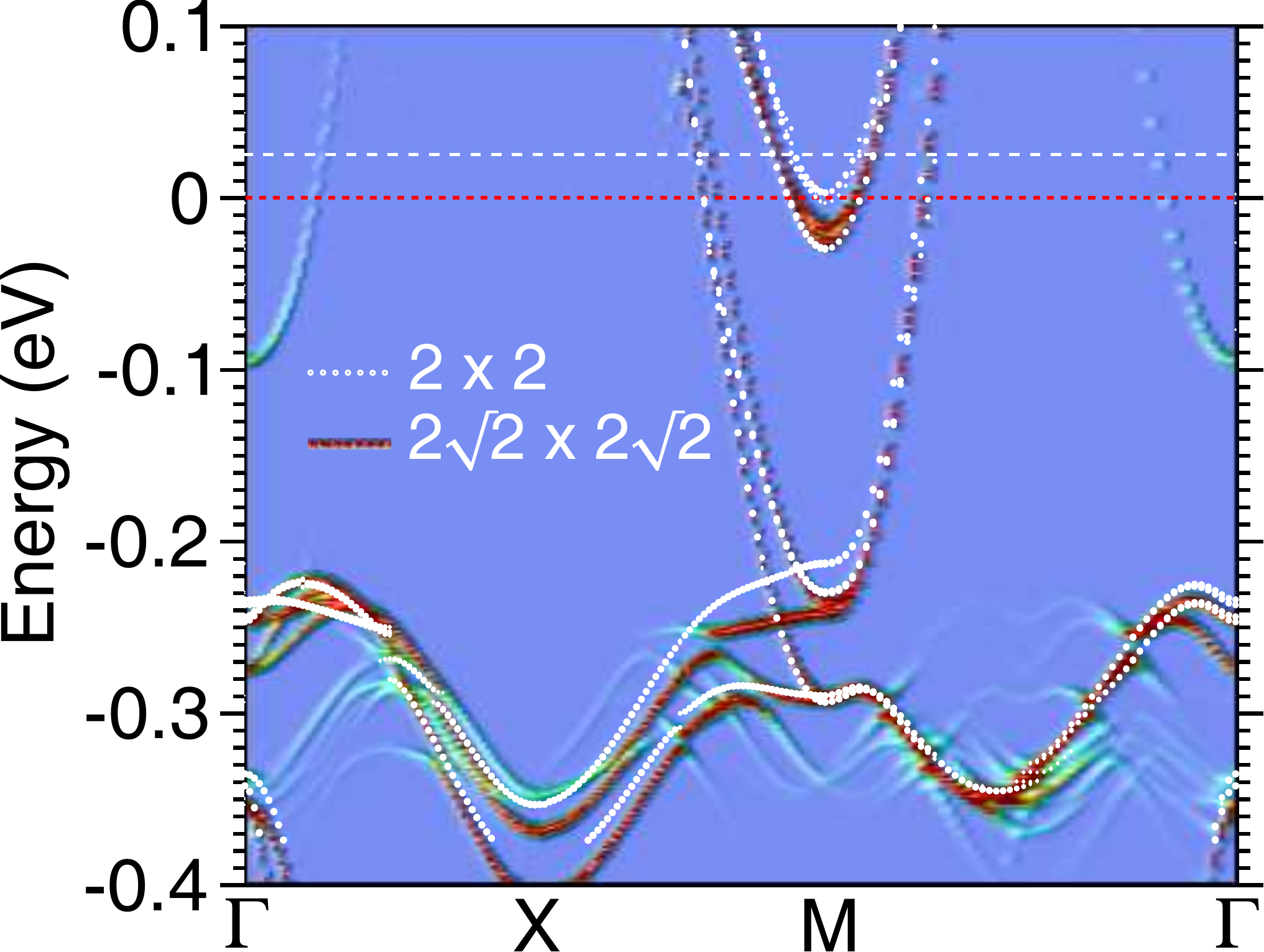}
  \caption{(color online) $k$-projected bands for CB-AFM FeSe monolayers on
SrTiO$_3$(001) with one O-vac in a $2\sqrt{2}\times2\sqrt{2}$ supercell.
The band structure for $2\times 2$ O-vacant FeSe/SrTiO$_3$(001) (white
curves, corresponding to the peaks of the $k$-projected FeSe bands in Fig.~\ref{fig2}e,
i.e., removing bands with low weight)
is overlaid for comparison.
The red (white) dashed line represent the Fermi level of
$2\sqrt{2}\times2\sqrt{2}$ ($2\times 2$) O-vacant FeSe/SrTiO$_3$(001).}
 \label{fig5}
\end{figure}

\begin{figure*}
  \includegraphics[width=0.90\textwidth]{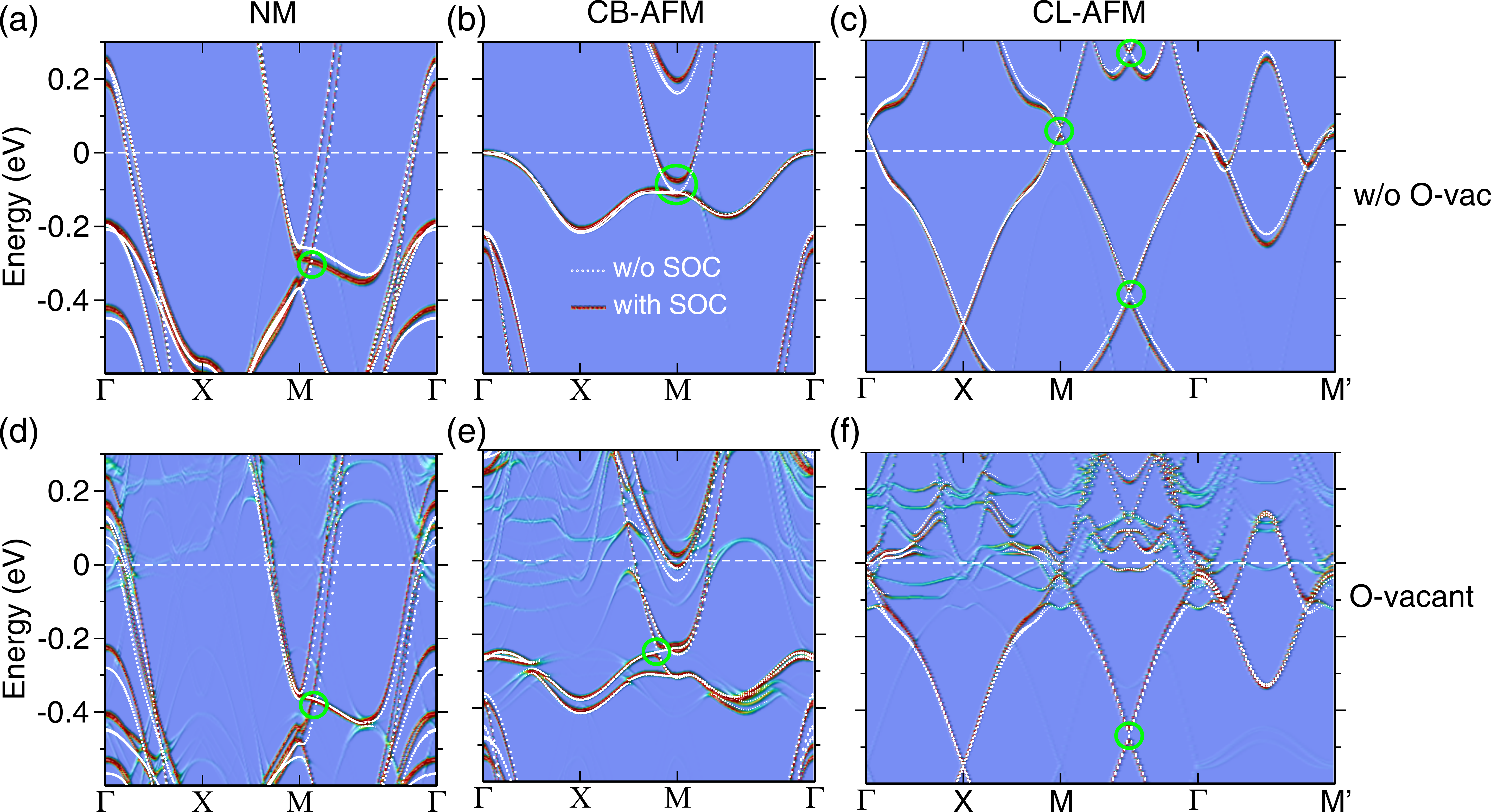}
  \caption{(color online) $k$-projected bands with SOC for (a)-(c) ideal
and (d)-(f) O-vacant FeSe/SrTiO$_3$(001).  Bands without SOC (white
curves) are overlaid for comparison.  Green circles mark SOC-induced
band splittings where bands are (nearly) degenerate without SOC.
  }
 \label{fig6}
\end{figure*}

Because the bands near $\Gamma$ and M for the CB-AFM state are sensitive
to oxygen vacancies, calculations were also performed for
$2\sqrt{2}\times2\sqrt{2}$ O-vacant FeSe/SrTiO$_3$(001) to see how the
bands change as the concentration of O-vac varies.  The substrate was
modeled by a single SrTiO$_3$ layer for computational reasons.  In this
case, the concentration of O-vac is half of that for $2\times 2$
O-vacant FeSe/SrTiO$_3$(001), which gives rise to less charge transfer
between FeSe and SrTiO$_3$(001).  Fig.~\ref{fig5} shows that the overall
band structure changes slightly.  The Fermi level shifts to a slightly
lower energy (Fig.~\ref{fig5}), consistent with a lower O-vac
concentration.  The gap at M located at about -0.25 eV decreases
slightly with lower of O-vac concentration.

\subsection{Spin-orbit coupling}

Spin-orbit coupling has been shown to affect the band structure in the
Fe-based superconductors \cite{soc-2015,borisenko_direct_2016}.  In
particular, it lifts the degeneracy at the zone center, inducing a band
splitting of $\sim$50 meV for FeTe$_{0.5}$Se$_{0.5}$.\cite{soc-2015}
The effects of SOC on the bands of FeSe films on SrTiO$_3$, both with
and without the inter-facial O-vac, are summarized in Fig.~\ref{fig6}.
One can see that SOC has noticeable effects on the topology of
electronic bands.  It induces band splittings (green circles) at/near
the M point where there is a band-crossing in the non-SOC calculations.
For the NM state SOC induces a splitting of about 60 meV for bands about
0.2 eV above the Fermi level at $\Gamma$ for the perfect
FeSe/SrTiO$_3$(001) (Fig.~\ref{fig6}(a)) and reduces the gap at M.  For
the CB-AFM configuration, SOC enhances the gap at M near E$_F$ by
$\sim$30 meV (Fig.~\ref{fig6}(b)); similar splittings of about 30 meV
are induced for the CL-AFM state.

\begin{figure}[b]
  \includegraphics[width=0.30\textwidth]{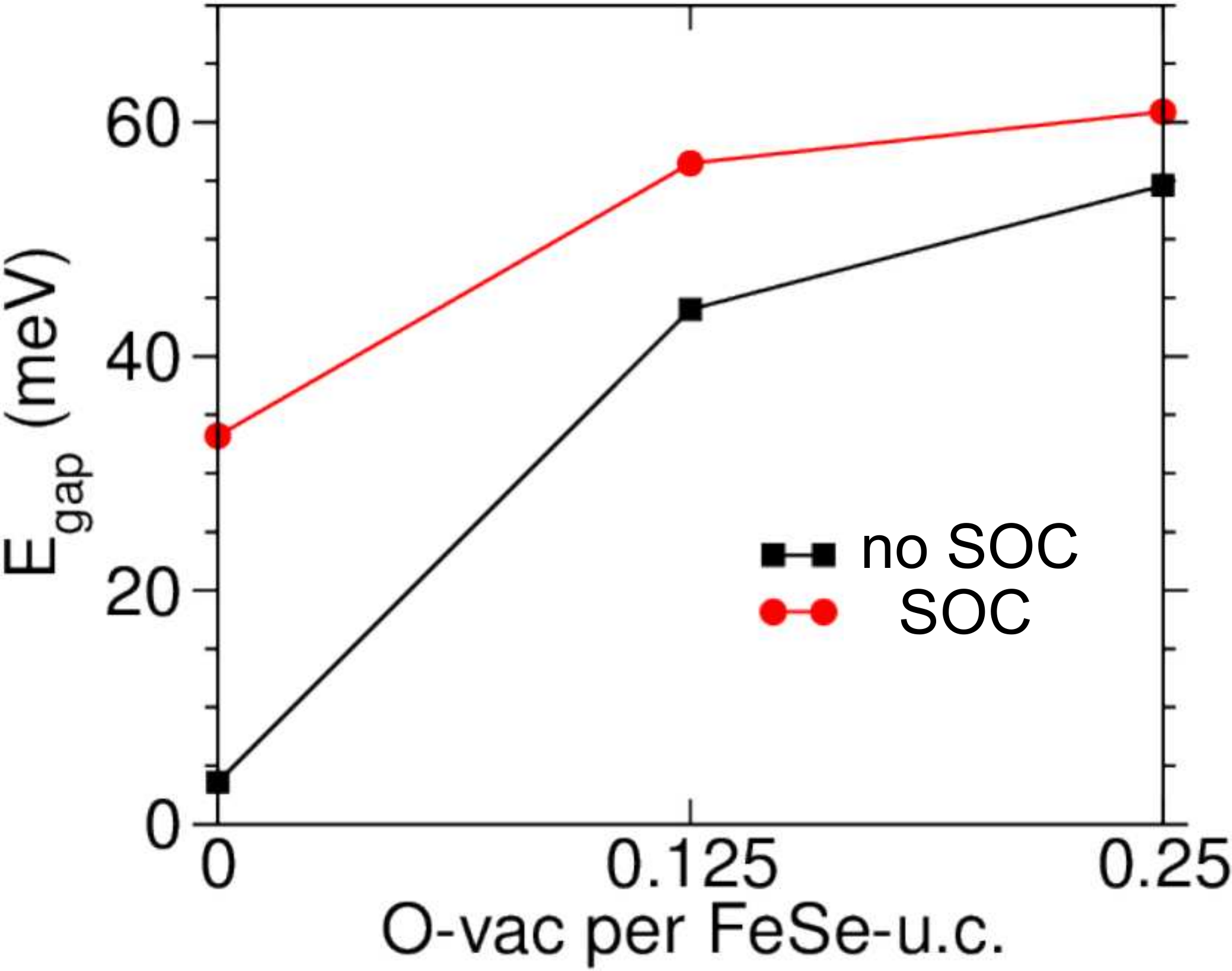}
  \caption{(color online) Effect of O-vac concentration on the gap near
the Fermi level at M for the CB-AFM state.  0.125 (0.25) O-vac/FeSe-u.c.
corresponds to the $2\sqrt{2}$$\times 2$$\sqrt{2}$ ($2$$\times$$2$)
O-vacant FeSe/SrTiO$_3$(001).
  }
 \label{fig7}
\end{figure}

The interfacial oxygen vacancy, which also caused band splittings, will
modify these splittings.  For the NM state,  a comparison of
Figs.~\ref{fig6} (a) and (d) shows that the presence of the interfacial
O-vac enhances the SOC-induced shift of the band near -0.3 eV at
$\Gamma$ by $\sim$ 30 meV.  For the CL-AFM, the O-vac has little effects
on SOC splittings, while for CB-AFM,  Fig.~\ref{fig6}(e), the gap at M
is enhanced by only about 5 meV.  Figure \ref{fig7} further depicts the
dependence of the SOC splitting on the O-vac concentration for the
CB-AFM state, and shows that the effects of SOC and the oxygen vacancies
are not simply additive.  This observation should not be surprising
since even in perturbation theory starting from states that are already
split (either by SOC or O-vacancies), adding in the other effect will be
suppressed because of the energy denominator. Note, however, that the
SOC contribution will cause a splitting for all vacancy contributions,
while the additional splitting attributable to the oxygen vacancies will
scale (in a complicated manner) with the vacancy concentration. There
is, however, a distinction between the SOC- and O vacancy-induced
splittings mainly along X-M, and to a lesser extent along $\Gamma$-X.

\subsection{Bands around $\Gamma$ and M}

\begin{figure}
  \includegraphics[width=0.35\textwidth]{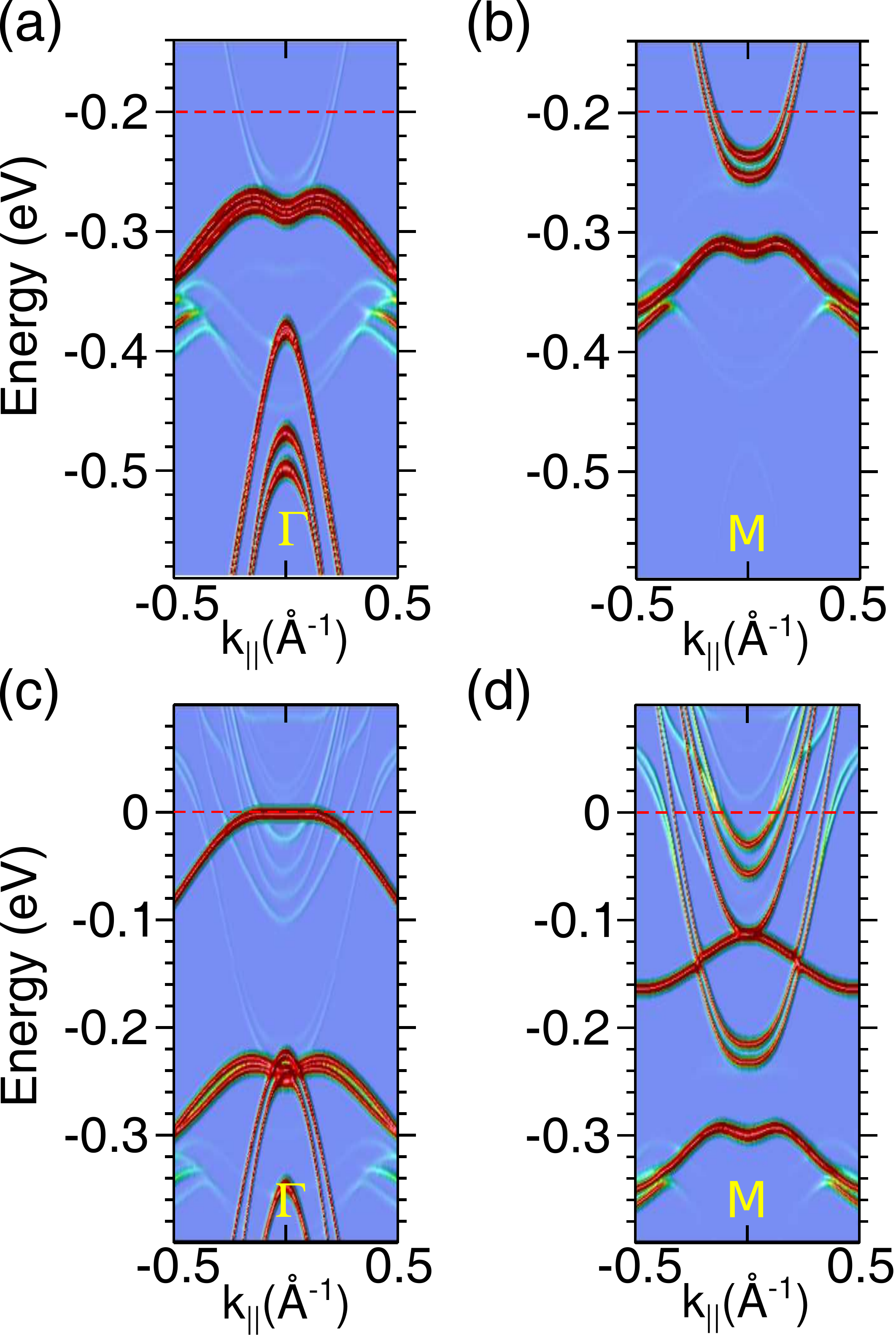}
  \caption{(color online) Bands for FeSe monolayer on STO-vac (a) around
$\Gamma$ (along M-$\Gamma$-M) and (b) M (along $\Gamma$-M-$\Gamma$), and
the (c) and (d) the corresponding bands for an FeSe bilayer.
}
 \label{fig10}
\end{figure}

Details of bands near $\Gamma$ and M for FeSe thin films on SrTiO$_3$
were investigated by ARPES experiments.
\cite{liu_electronic_2012,he_phase_2013,tan_2013,lee_interfacial_2014,liu_dichotomy_2014}
It was observed that the Fermi surface of monolayer FeSe/SrTiO$_3$
consists of electron-like pockets at M with a band bottom 60 meV below
the Fermi
level.\cite{liu_electronic_2012,he_phase_2013,tan_2013,liu_dichotomy_2014}
Further analysis reveals that there are two nearly degenerate electron
bands.\cite{lee_interfacial_2014} Moreover, there is one hole-like band
about 40 meV below the bottom of the electron-like band.  At $\Gamma$
there is one hole-like band with a top located 80 meV below E$_F$.  In
the case of bilayer FeSe/SrTiO$_3$ there are several hole-like bands
right below E$_F$ at $\Gamma$.  There are four small pockets forming a
cross-shaped Fermi surface centered around M.  Detailed band structure
reveals that they result from the crossing of a hole-like band and a
electron-like band.  At the M point there is another hole-like band
below the Fermi level with a top at 60 meV below E$_F$ which has a
energy separation of about 80 meV from the higher hole band.  The
separation/splitting reduces dramatically as the $k$ point goes further
away from M (Fig.~S2 in Ref.\ \onlinecite{tan_2013}).

Our calculations show that both the NM state and the CL-AFM state give
rise to large pockets around $\Gamma$, inconsistent with the
experimental observations.  In contrast, for the CB-AFM state
(Fig.~\ref{fig2}(e)) there is no pocket near $\Gamma$ and only electron
pockets centered about M, consistent with the ARPES results.  Since in
the present study a 2$\times$2 supercell with one O-vacancy was used to
model the substrate, this gives rise to a much  higher doping level than
experiment and results in a larger shift of the Fermi level.  Based on
these considerations, we rigidly shift down the Fermi level such that it
corresponds to a lower doping level ($\sim$0.2 $e^-$/Fe); 
based on the 2$\times$2 and $\sqrt{2}$$\times$$\sqrt{2}$ supercells, 
a single oxygen vacancy dopes $\sim$1.6--1.7 electrons into the FeSe overlayer.
Figs.~\ref{fig10}(a) and (b) show
bands about $\Gamma$ and M, respectively.  One can see that there are
two nearly degenerate electron bands about M, where the band splitting
between the electron band and the hole band is $\sim$40 meV, consistent
with the ARPES experiments.  In addition, the width of the bands of
interest about $\Gamma$ and M are also in agreement with ARPES results
(Fig.~3 in Ref.\ \onlinecite{liu_dichotomy_2014}) without any rescaling.
The shape of the top of the hole-band at $\Gamma$ as well as M is
sensitive to lattice constant, which becomes round for smaller in-plane
lattice constants.  The replica bands seen by the ARPES
experiments\cite{lee_interfacial_2014} does not appear in our
calculations, these have been attributed to the coupling of FeSe bands
to the substrate ferroelectric phonons.

For bilayer FeSe/SrTiO$_3$(001), the Fermi level of the top FeSe layer
stays nearly unchanged in the presence of the O-vacancy.
Figure~\ref{fig10}(c) shows that there is only one hole-like band at
$\Gamma$ near E$_F$, differing from the ARPES measurements which observe
multiple hole bands.  In Fig.~\ref{fig10}(d), the splitting of about 180
meV between the two hole-like bands at the M point is much larger than
that is seen by the experiments, and the splitting tends to increase as
the $k$ moves away from M, in contrast to the experimental observations.
Shifting the bands of the interface FeSe layer to account for a lower
electron doping and then superimposing them onto those for the top FeSe
layer does not improve the agreement.  Given that the 80 meV splitting
of the two hole bands at M are common for FeSe thin films thicker than
monolayer,\cite{tan_2013} the discrepancy between our calculations and
ARPES measurements is unclear, but might be attributed to nematicity in
FeSe, which is not directly taken into account in our calculations.

\subsection{Se-O (Type B) interface}

\begin{figure}
 \includegraphics[width=0.48\textwidth]{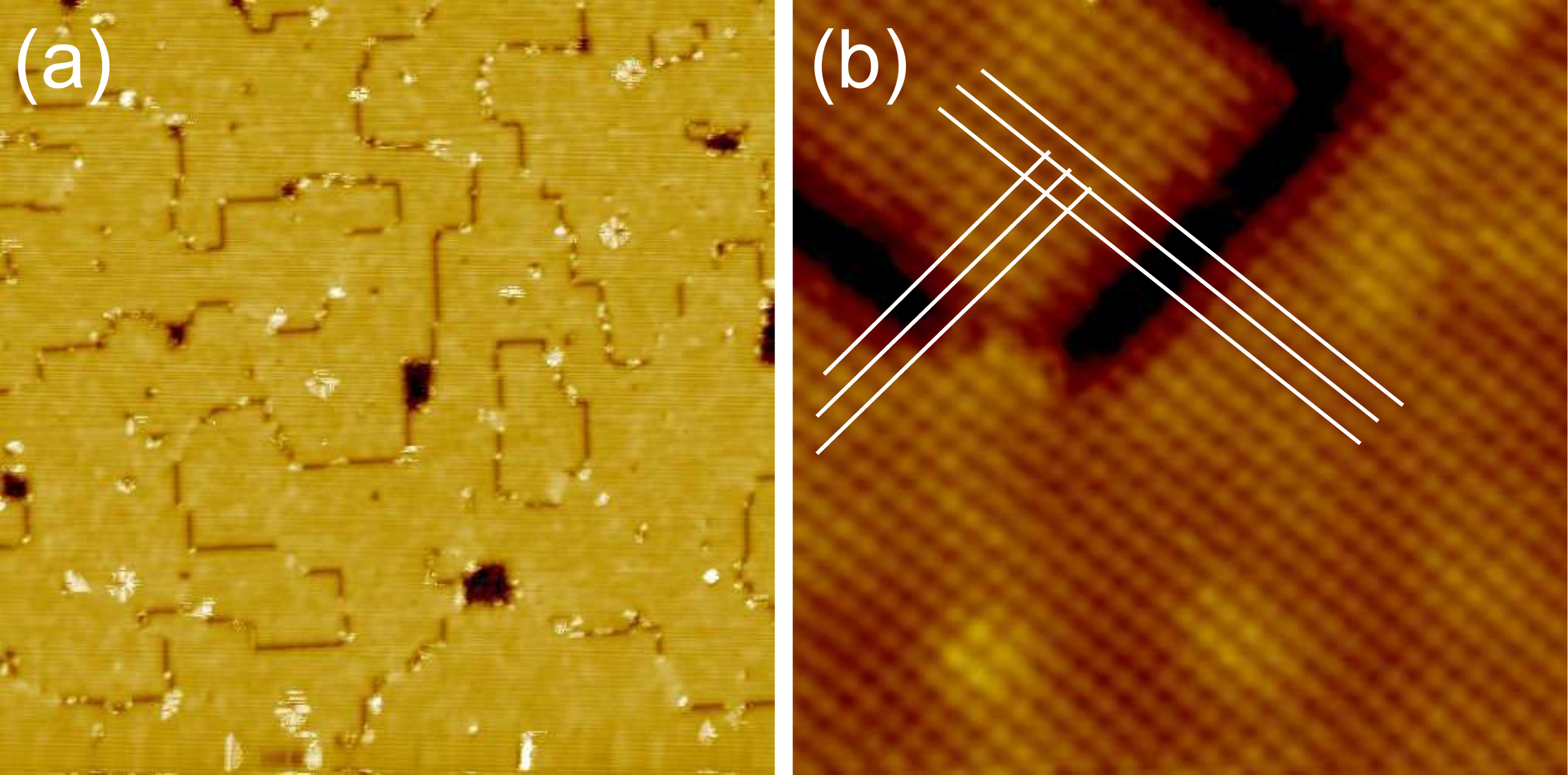}
 \caption{(color online)
(a) STM image of a MBE-grown FeSe monolayer film on SrTiO$_3$(001)
 (V$_\textrm{s}$ = 1.5 V, I$_\textrm{t}$ = 0.2 nA, 100 nm $\times$ 100 nm).
(b) Close-up view of a grain boundary (V$_\textrm{s}$ = $-$0.5 V,
I$_\textrm{t}$ = 1.5 nA, 10 nm $\times$ 10 nm). The lines are guides to
show the relative shift in alignment between the two grains.
}
\label{fig-expt1}
\end{figure}

Up to now, we have focused on the Type A, Se-Ti, epitaxial relationship
since this is the preferred for the ideal interface. The overall
energetics do not change in the presence of an oxygen vacancy: for
example, our calculations for 2$\sqrt{2}$$\times$2$\sqrt{2}$ O-vacancy
FeSe/SrTiO$_3$(001) supercells, Type A (Se-Ti) is 0.25 eV/FeSe-u.c.
lower than the Type B (Se-O), in large part because of the increase in
the interlayer distance to about 3.34 \AA. Despite these energetics, if
the oxygen vacancy were to nucleate a Type B registry of the FeSe film,
the barrier to shift may be large enough to pin the film.

Experimentally, there is evidence for different possible epitaxial
relationships.  
The growth of FeSe films was carried out in an integrated MBE-STM
ultrahigh vacuum system with a base pressure of 2$\times$10$^{-11}$
Torr. Monolayer and bilayer FeSe films were grown on 0.05 wt\%
Nb-doped SiTiO$_3$(001) substrates, which were first annealed at
900 $^\circ$C in Se flux. Then the FeSe films were grown under Se-rich
conditions with a growth rate of 0.2 monolayers per minute. Monolayer
FeSe films were further annealed at $\sim$500 $^\circ$C for 2-3 hours to
remove adsorbed Se and to reach a superconducting state with a paring
gap of $\sim$20 meV, consistent with the earlier studies.\cite{qing-yan_2012}
Scanning tunneling
microscopy (STM) images,
Fig.~\ref{fig-expt1}, show that the
films consist of large grains, with their crystallographic axes
aligned, presumably with the SrTiO$_3$. However, as shown in the atomic
resolution image in Fig.~\ref{fig-expt1}(b), the films across a grain
boundary can be shifted relative to each, i.e., the FeSe films may have
various epitaxial relationships to the SrTiO$_3$ substrate.

\begin{figure}
  \includegraphics[width=0.48\textwidth]{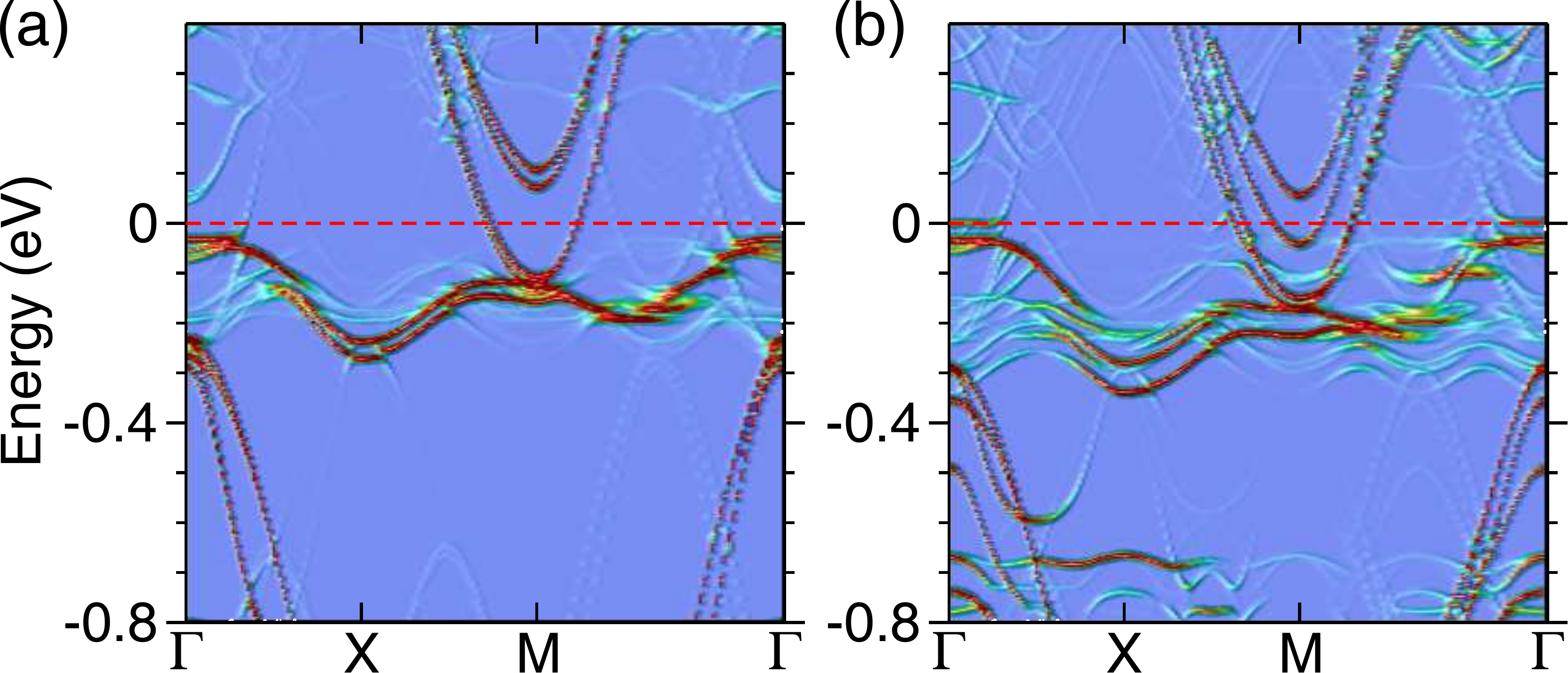}
  \caption{(color online) Electronic bands for the CB-AFM FeSe in Type B
registry (Se above O) for $2\sqrt{2}\times2\sqrt{2}$ O-vacancy
FeSe/SrTiO$_3$(001) at (a) the equilibrium and (b) reduced (by 0.8 \AA)
interlayer separation.
  }
 \label{fig9}
\end{figure}

The unfolded bands for the CB-AFM Type B interface (with O vacancy) is
shown in Fig.~\ref{fig9}(a).  Comparing to Fig.~\ref{fig5} shows that
there is only a small shift of the Fermi level, implying less charge
transferred to the FeSe layer for this configuration, due in part to the
increased separation.  Artificially reducing the layer distance by about
0.8 \AA{}, Fig.~\ref{fig9}(b), leads to a strong hybridization between
the FeSe monolayer and the substrate (Fig.~\ref{fig9}(b)), causing
noticeable changes around the M point.

\subsection{Se-O vacancy binding}

\begin{figure}[b]
  \includegraphics[width=0.48\textwidth]{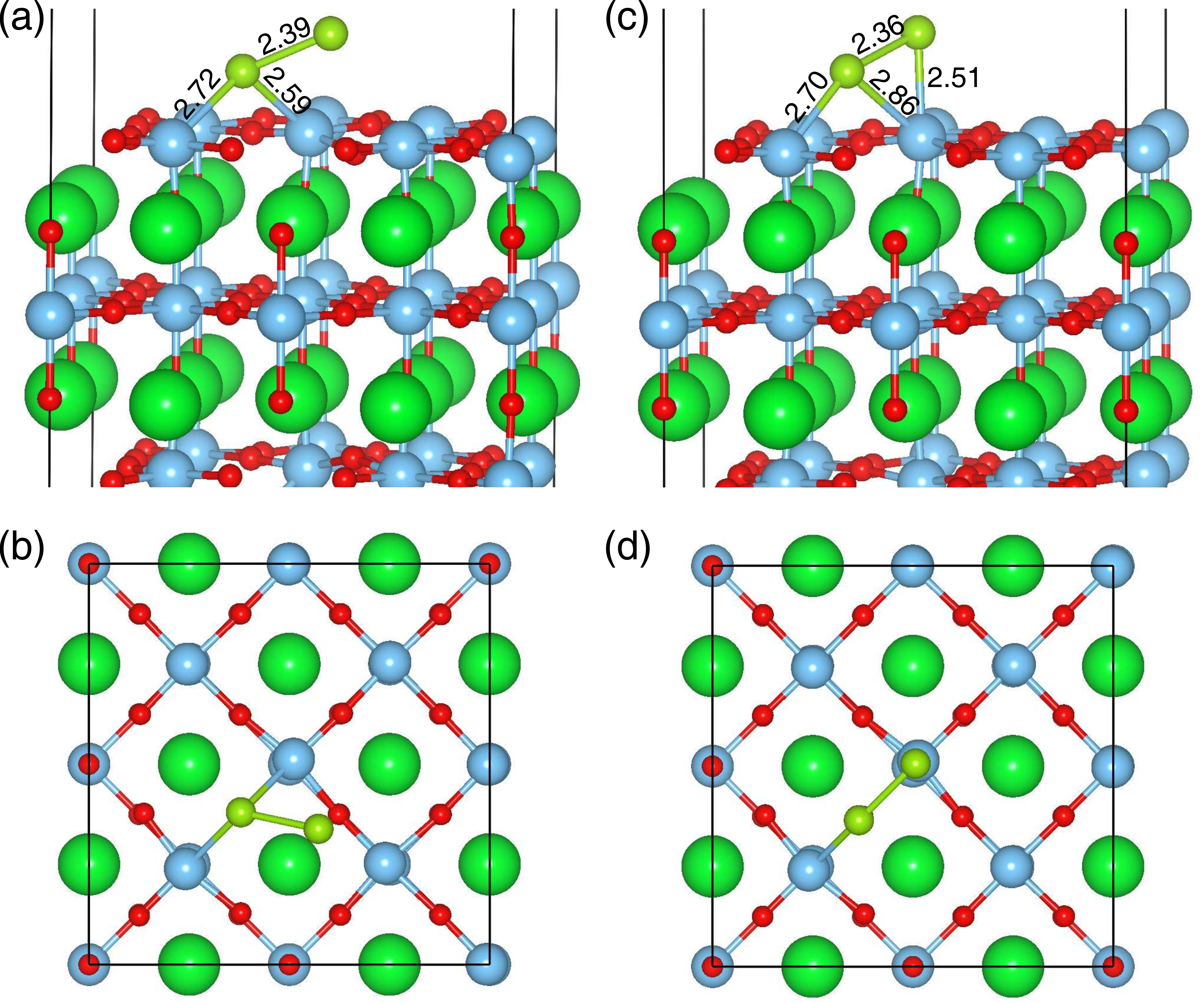}
  \caption{(color online) Calculated structure of two Se deposited onto
SrTiO$_3$(001) with one O vacancy in a $2\sqrt{2}\times2\sqrt{2}$
supercell.  The two lowest-energy configurations out of the considered
candidates are shown:
  (a) Perspective and (b) top views of configuration S1, 
  obtained from the structural relaxation 
  starting with the second Se sitting on an oxygen atom close to the O-vac;
  (c) and (d) same for S2.
  Bond lengths involving Se are shown (in \AA).
  }
 \label{fig8}
\end{figure}

Although oxygen vacancies are expected surface defects of SrTiO$_3$, it
is not obvious that these will be the dominant defects in the combined
system since  experimentally the surfaces of SrTiO$_3$ are treated by Se
flux before the growth of FeSe thin films.\cite{qing-yan_2012} This
raises the possibility that Se interacts with the oxygen vacancies. Our
calculations show that a single Se has a strong tendency to bind to an O
vacancy: A single Se on SrTiO$_3$(001) with one O-vac in a
$2\sqrt{2}\times2\sqrt{2}$ supercell prefers the binding to the O
vacancy rather than to a Ti by 1.20 eV, i.e., Se have a strong driving
force to saturate O vacancies.  The bond lengths of Se and its
neighboring Ti atoms are about 2.52 \AA.  Introducing another Se
originally sitting on the top site of the O near the O-vac leads to the
formation of Se dimer of Se after structural relaxation (hereafter
referred as S1, Figs.~\ref{fig8}(a) and (b)). The Se binding to the
O-vac moves further away from the SrTiO$_3$(001) surface such that one
Se-Ti bond is increased up to about 2.70 \AA.  The second Se is further
away from the SrTiO$_3$(001) surface by about 1.02 \AA{} than the one
sitting on the O-vac.  Another configuration (S2, Figs.~\ref{fig8}(c)
and (d)) with the second Se originally placed on the top site of Ti near
the O-vac is only 0.06 eV/Se higher than S1, indicating the tendency of
Ti as the preferable sites for Se when the O-vac is saturated. Thus, if
Se atoms can saturate the oxygen vacancies, then a Type A interface may
result.

\begin{figure}
  \includegraphics[width=0.45\textwidth]{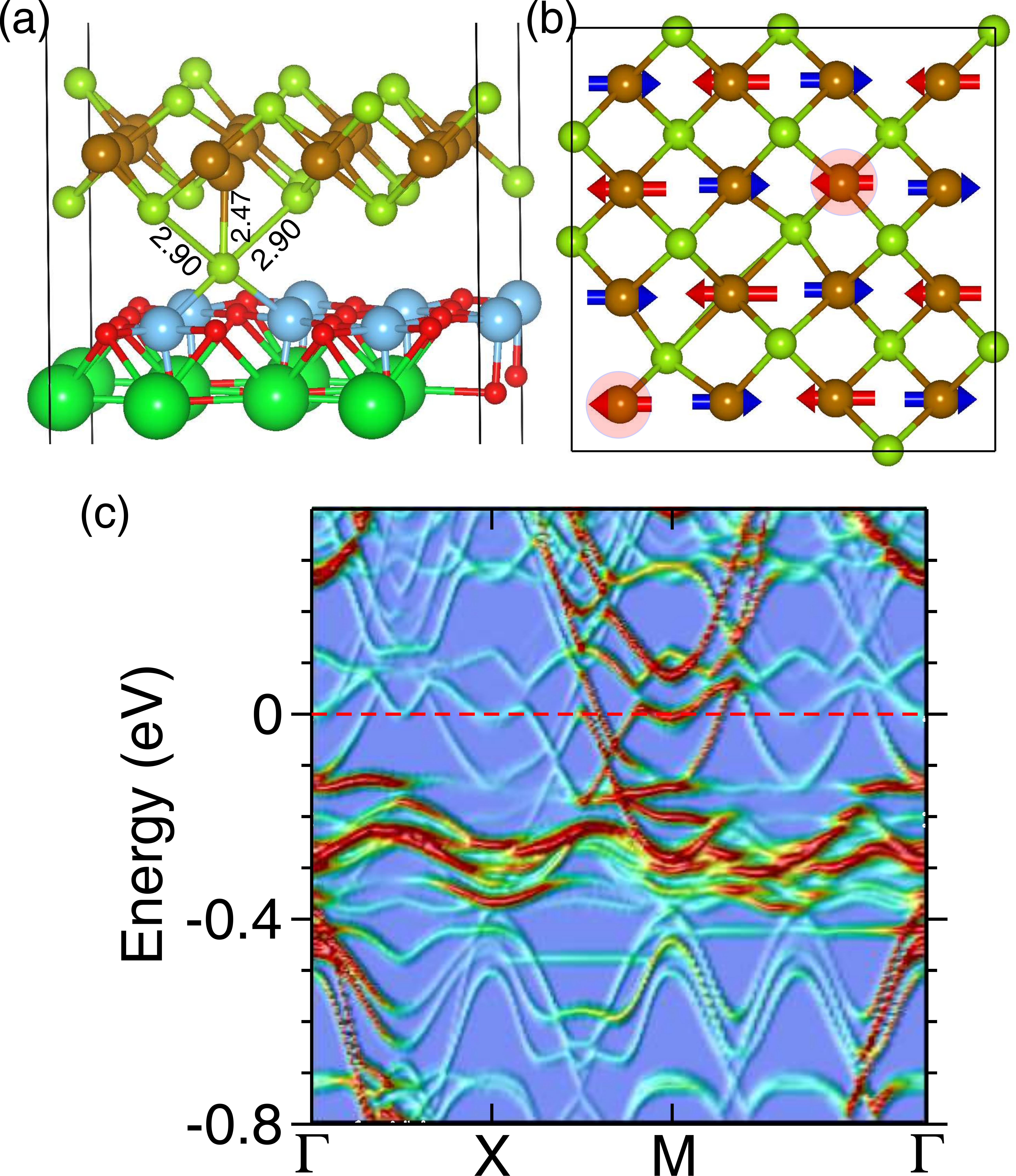}
  \caption{(color online) Se-O vacancy Type A  FeSe/SrTiO$_3$(001) interface: 
  (a) Perspective view of the relaxed structure for CB-AFM
$2\sqrt{2}\times2\sqrt{2}$ cell, and
  (b) schematic of the magnetic structure (shaded atoms indicate that
the moments on them are significantly reduced).
  (c) $k$-projected band structure. 
  }
 \label{fig11}
\end{figure}

Because Se is isovalent to O, filling the O-vac by Se may be expected
gives rise to similar geometric and electronic properties as the perfect
FeSe/SrTiO$_3$(001).  However, our calculations reveal that the filled
Se induces dramatic changes in the structure of the FeSe monolayer.  The
calculation was carried out for the CB-AFM $2\sqrt{2}\times2\sqrt{2}$
FeSe/SrTiO$_3$(001) where the two constituents are in Se-Ti (Type A)
stacking.  The relaxed structure is shown in Fig.~\ref{fig11}(a).  The
extra Se forms direct bonding with the Fe sitting above it, which pulls
the Fe out (down) of the Fe plane by about 0.55 \AA.  The two surface Se
atoms binding to the downward-shifted Fe are correspondingly shifted,
that is, they are pulled closer to the Fe plane by about 0.25 \AA.

\begin{figure}
  \includegraphics[width=0.48\textwidth]{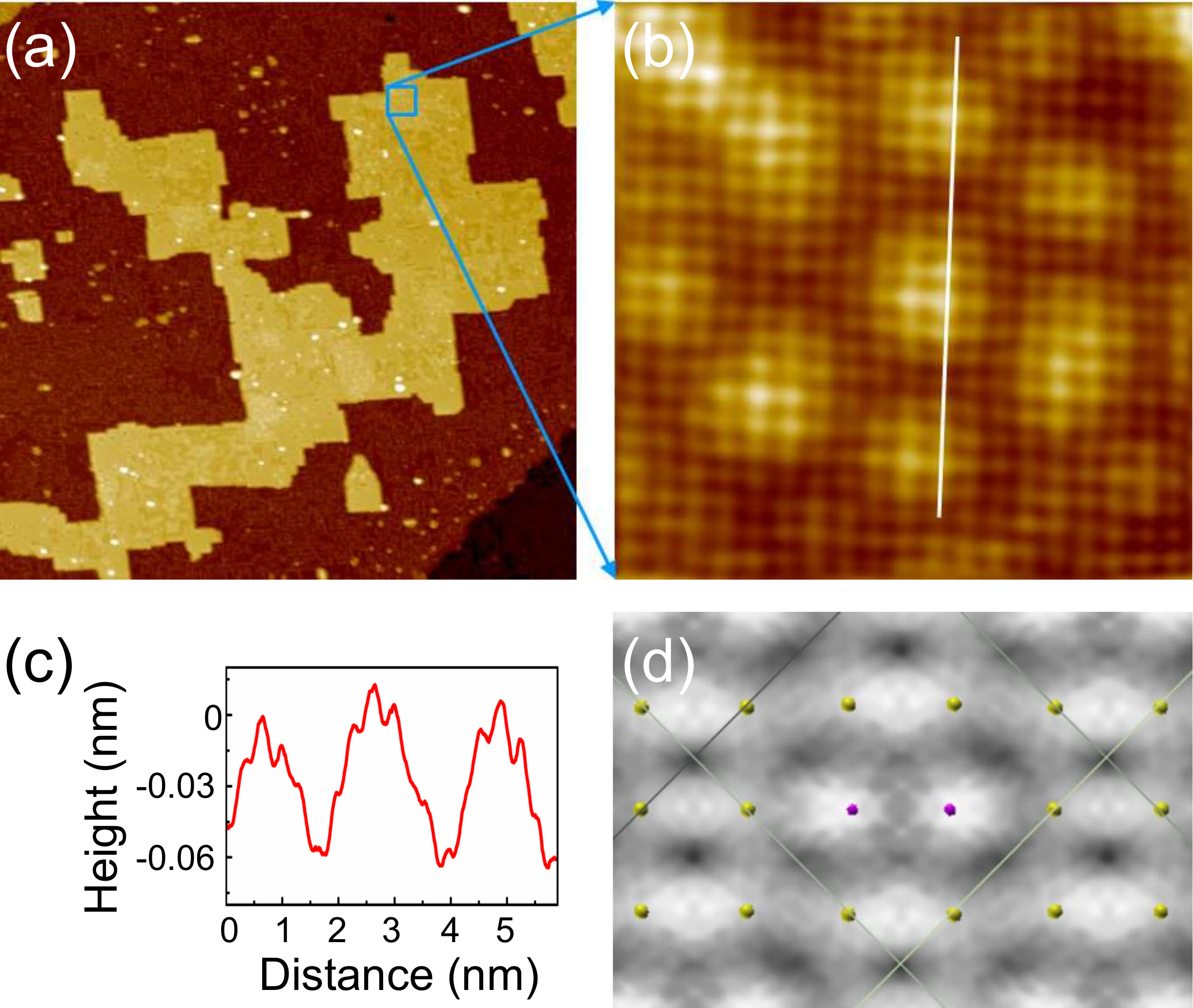}
  \caption{(color online)
  (a) STM image showing monolayer FeSe islands on SrTiO$_3$(001)
      (V$_\textrm{s}$ = 1.0 V, I$_\textrm{t}$ = 0.1 nA, 200 nm $\times$ 200 nm).
  (b) Atomic resolution image of the boxed region in (a)
      (V$_\textrm{s}$ = 1.0 V, I$_\textrm{t}$ = 0.1 nA, 10 nm $\times$ 10 nm).
  (c) Line profile (height variations) along the line shown in (b).
  (d) Calculated STM image for an FeSe monolayer on the Se--O-vacancy complex.
The positions of the uppermost Se atoms are indicated; the purple dots indicate the Se
atoms closest to the defect, which are pulled down by $\sim$0.25 \AA{} relative to the
other Se atoms. (Constant current/density simulation: 1 V bias,
2$\times$10$^{-8}$$e$/\AA$^3$; nominal average height of 4.6 \AA{} above the surface,
with a corrugation of $\pm$0.06 \AA).
}
 \label{fig-expt2}
\end{figure}

Defects in MBE-grown FeSe films are not unusual. A common defect appears
as enhanced contrast over $\sim$4 unit cells, with a height fluctuation
of several hundredths of a nm, c.f., Figs.~\ref{fig-expt2}(a,b) and is seen throughout
the sample.
Although direct comparisons between the experiments and calculations are
not feasible because of much larger supercells required, our results are
at least suggestive. In contrast to the oxygen vacancy case in which the
interlayer separation between the FeSe and substrate decreases, for the
Se--O-vacancy complex there is an increase of 0.1--0.2 \AA. Although low
compared to the values in Fig.~\ref{fig-expt2}(c) (and 
closer to the values calculated for the Type B interface), there is not
a simple one-to-one relationship between the STM height differences and
the atom positions. In fact, the simulated STM image,
Fig.~\ref{fig-expt2}(d), shows enhanced corrugation at the Se atoms
nearest the defect complex, despite the fact that these are 0.25 \AA{} lower. We note
that simulated STM images (not shown) for the simple O vacancy and the Type B interface
appear less consistent with the experimental data. While the present results do not
conclusively demonstrate that Se--O-vacancy complexes exist, they do strongly suggest
that defect complexes -- not just simple oxygen vacancies -- exist at the
substrate-FeSe interface and can modify the electronic properties.

This conclusion is borne out in the calculations for FeSe at the
Se--O-vacancy interface. There is no net magnetic moment for the ideal
free-standing AFM FeSe monolayer, but in proximity to either the ideal
or the O-vacant SrTiO$_3$(001), small ($<$0.06 $\mu_B$)  net magnetic
moments are induced due to inversion-symmetry breaking.  When the Se
fills the O-vacancy, changes in the Fe magnetic moments 
give rise to a net magnetic moment of about 1.0 $\mu_B$ for the FeSe
overlayer.  The moment on the downward-shifted Fe is significantly
enhanced by about 0.5 $\mu_B$, and a general increase in that spin
channel, except for the two Fe atoms marked in Fig.~\ref{fig11}(b),
which are reduced by about 0.3 $\mu_B$.  The dramatic change in the
magnetism has an important effect on the band structure of the FeSe
monolayer.  Fig.~\ref{fig11}(c) shows that the electronic bands of FeSe
are strongly disturbed by the fluctuating magnetism, though the global
profile is similar to that for CB-AFM FeSe monolayer.  A comparison of
Fig.~\ref{fig11}(c) and Fig.~\ref{fig5} shows that the size of the
electron pockets centered at M decreases, suggesting that the extra Se
lowers the doping level of the FeSe overlayer.  It should be noted that
the origin of the doping to the FeSe monolayer on Se-filled
SrTiO$_3$(001) is different from that for O-vacant FeSe/SrTiO$_3$(001);
in the former case, the doping is due to the charge transfer between the
extra Se and the FeSe overlayer, whereas, in the latter the O-vac is
mainly responsible for the doping.

\section{Summary}

In summary, we have investigated the effects of oxygen vacancies at the
interface of between FeSe thin films and SrTiO$_3$(001).  The interface
oxygen vacancy not only provides electron doping to the interface FeSe
layer, but also significantly renormalizes the width of the Fe-3$d$ band
near the Fermi level for the checkerboard AFM state.  However, due to
the screening of the interface FeSe layer, the effects of the O-vacancy
on the electronic properties of the top FeSe layer in bilayer FeSe are
limited, e.g., the electronic bands for the top FeSe layer are similar
to the perfect monolayer FeSe/SrTiO$_3$(001).  Spin-orbit is also found
to play an important role in determining the detailed band structure,
giving rise to splittings both for the ideal case and in the presence of
defects. Our results suggest that while different epitaxial
relationships are possible in the presence of simple oxygen vacancies, a
likely defect is a Se--O-vacancy, with then ``standard'' Type A (Se-Ti)
registry.  The calculated bands around $\Gamma$ and M for the
checkerboard-AFM monolayer FeSe/SrTiO$_3$(001) in the presence of oxygen
vacancies are generally consistent with ARPES results.  However, for the
bilayer case the agreement is not particularly good, which requires
further investigations.  Finally, our results demonstrate that
understanding the detailed electronic properties requires treating a
number of different effects on an equal footing: short- and long-range
magnetic correlations, spin-orbit, epitaxial relationships, and defect
complexes.

\begin{acknowledgments}
This work was supported by the U.S. National Science Foundation,
Division of Materials Research, DMR-1335215.
\end {acknowledgments}

\bibliography{references}

\begin{thebibliography}{40}%
\makeatletter
\providecommand \@ifxundefined [1]{%
 \@ifx{#1\undefined}
}%
\providecommand \@ifnum [1]{%
 \ifnum #1\expandafter \@firstoftwo
 \else \expandafter \@secondoftwo
 \fi
}%
\providecommand \@ifx [1]{%
 \ifx #1\expandafter \@firstoftwo
 \else \expandafter \@secondoftwo
 \fi
}%
\providecommand \natexlab [1]{#1}%
\providecommand \enquote  [1]{``#1''}%
\providecommand \bibnamefont  [1]{#1}%
\providecommand \bibfnamefont [1]{#1}%
\providecommand \citenamefont [1]{#1}%
\providecommand \href@noop [0]{\@secondoftwo}%
\providecommand \href [0]{\begingroup \@sanitize@url \@href}%
\providecommand \@href[1]{\@@startlink{#1}\@@href}%
\providecommand \@@href[1]{\endgroup#1\@@endlink}%
\providecommand \@sanitize@url [0]{\catcode `\\12\catcode `\$12\catcode
  `\&12\catcode `\#12\catcode `\^12\catcode `\_12\catcode `\%12\relax}%
\providecommand \@@startlink[1]{}%
\providecommand \@@endlink[0]{}%
\providecommand \url  [0]{\begingroup\@sanitize@url \@url }%
\providecommand \@url [1]{\endgroup\@href {#1}{\urlprefix }}%
\providecommand \urlprefix  [0]{URL }%
\providecommand \Eprint [0]{\href }%
\providecommand \doibase [0]{http://dx.doi.org/}%
\providecommand \selectlanguage [0]{\@gobble}%
\providecommand \bibinfo  [0]{\@secondoftwo}%
\providecommand \bibfield  [0]{\@secondoftwo}%
\providecommand \translation [1]{[#1]}%
\providecommand \BibitemOpen [0]{}%
\providecommand \bibitemStop [0]{}%
\providecommand \bibitemNoStop [0]{.\EOS\space}%
\providecommand \EOS [0]{\spacefactor3000\relax}%
\providecommand \BibitemShut  [1]{\csname bibitem#1\endcsname}%
\let\auto@bib@innerbib\@empty
\bibitem [{\citenamefont {Qing-Yan}\ \emph {et~al.}(2012)\citenamefont
  {Qing-Yan}, \citenamefont {Zhi}, \citenamefont {Wen-Hao}, \citenamefont
  {Zuo-Cheng}, \citenamefont {Jin-Song}, \citenamefont {Wei}, \citenamefont
  {Hao}, \citenamefont {Yun-Bo}, \citenamefont {Peng}, \citenamefont {Kai},
  \citenamefont {Jing}, \citenamefont {Can-Li}, \citenamefont {Ke},
  \citenamefont {Jin-Feng}, \citenamefont {Shuai-Hua}, \citenamefont {Ya-Yu},
  \citenamefont {Li-Li}, \citenamefont {Xi}, \citenamefont {Xu-Cun},\ and\
  \citenamefont {Qi-Kun}}]{qing-yan_2012}%
  \BibitemOpen
  \bibfield  {author} {\bibinfo {author} {\bibfnamefont {W.}~\bibnamefont
  {Qing-Yan}}, \bibinfo {author} {\bibfnamefont {L.}~\bibnamefont {Zhi}},
  \bibinfo {author} {\bibfnamefont {Z.}~\bibnamefont {Wen-Hao}}, \bibinfo
  {author} {\bibfnamefont {Z.}~\bibnamefont {Zuo-Cheng}}, \bibinfo {author}
  {\bibfnamefont {Z.}~\bibnamefont {Jin-Song}}, \bibinfo {author}
  {\bibfnamefont {L.}~\bibnamefont {Wei}}, \bibinfo {author} {\bibfnamefont
  {D.}~\bibnamefont {Hao}}, \bibinfo {author} {\bibfnamefont {O.}~\bibnamefont
  {Yun-Bo}}, \bibinfo {author} {\bibfnamefont {D.}~\bibnamefont {Peng}},
  \bibinfo {author} {\bibfnamefont {C.}~\bibnamefont {Kai}}, \bibinfo {author}
  {\bibfnamefont {W.}~\bibnamefont {Jing}}, \bibinfo {author} {\bibfnamefont
  {S.}~\bibnamefont {Can-Li}}, \bibinfo {author} {\bibfnamefont
  {H.}~\bibnamefont {Ke}}, \bibinfo {author} {\bibfnamefont {J.}~\bibnamefont
  {Jin-Feng}}, \bibinfo {author} {\bibfnamefont {J.}~\bibnamefont {Shuai-Hua}},
  \bibinfo {author} {\bibfnamefont {W.}~\bibnamefont {Ya-Yu}}, \bibinfo
  {author} {\bibfnamefont {W.}~\bibnamefont {Li-Li}}, \bibinfo {author}
  {\bibfnamefont {C.}~\bibnamefont {Xi}}, \bibinfo {author} {\bibfnamefont
  {M.}~\bibnamefont {Xu-Cun}}, \ and\ \bibinfo {author} {\bibfnamefont
  {X.}~\bibnamefont {Qi-Kun}},\ }\href {\doibase 10.1088/0256-307X/29/3/037402}
  {\bibfield  {journal} {\bibinfo  {journal} {Chin. Phys. Lett.}\ }\textbf
  {\bibinfo {volume} {29}},\ \bibinfo {pages} {037402} (\bibinfo {year}
  {2012})}\BibitemShut {NoStop}%
\bibitem [{\citenamefont {Huang}\ \emph {et~al.}(2015)\citenamefont {Huang},
  \citenamefont {Song}, \citenamefont {Webb}, \citenamefont {Fang},
  \citenamefont {Chang}, \citenamefont {Moodera}, \citenamefont {Kaxiras},\
  and\ \citenamefont {Hoffman}}]{huang_2015}%
  \BibitemOpen
  \bibfield  {author} {\bibinfo {author} {\bibfnamefont {D.}~\bibnamefont
  {Huang}}, \bibinfo {author} {\bibfnamefont {C.-L.}\ \bibnamefont {Song}},
  \bibinfo {author} {\bibfnamefont {T.~A.}\ \bibnamefont {Webb}}, \bibinfo
  {author} {\bibfnamefont {S.}~\bibnamefont {Fang}}, \bibinfo {author}
  {\bibfnamefont {C.-Z.}\ \bibnamefont {Chang}}, \bibinfo {author}
  {\bibfnamefont {J.~S.}\ \bibnamefont {Moodera}}, \bibinfo {author}
  {\bibfnamefont {E.}~\bibnamefont {Kaxiras}}, \ and\ \bibinfo {author}
  {\bibfnamefont {J.~E.}\ \bibnamefont {Hoffman}},\ }\href {\doibase
  10.1103/PhysRevLett.115.017002} {\bibfield  {journal} {\bibinfo  {journal}
  {Phys. Rev. Lett.}\ }\textbf {\bibinfo {volume} {115}},\ \bibinfo {pages}
  {017002} (\bibinfo {year} {2015})}\BibitemShut {NoStop}%
\bibitem [{\citenamefont {Zhou}\ \emph {et~al.}(2015)\citenamefont {Zhou},
  \citenamefont {Zhang}, \citenamefont {Liu}, \citenamefont {Tang},
  \citenamefont {Wang}, \citenamefont {Li}, \citenamefont {Song}, \citenamefont
  {Ji}, \citenamefont {He}, \citenamefont {Wang}, \citenamefont {Ma},\ and\
  \citenamefont {Xue}}]{zhou_2015}%
  \BibitemOpen
  \bibfield  {author} {\bibinfo {author} {\bibfnamefont {G.}~\bibnamefont
  {Zhou}}, \bibinfo {author} {\bibfnamefont {D.}~\bibnamefont {Zhang}},
  \bibinfo {author} {\bibfnamefont {C.}~\bibnamefont {Liu}}, \bibinfo {author}
  {\bibfnamefont {C.}~\bibnamefont {Tang}}, \bibinfo {author} {\bibfnamefont
  {X.}~\bibnamefont {Wang}}, \bibinfo {author} {\bibfnamefont {Z.}~\bibnamefont
  {Li}}, \bibinfo {author} {\bibfnamefont {C.}~\bibnamefont {Song}}, \bibinfo
  {author} {\bibfnamefont {S.}~\bibnamefont {Ji}}, \bibinfo {author}
  {\bibfnamefont {K.}~\bibnamefont {He}}, \bibinfo {author} {\bibfnamefont
  {L.}~\bibnamefont {Wang}}, \bibinfo {author} {\bibfnamefont {X.}~\bibnamefont
  {Ma}}, \ and\ \bibinfo {author} {\bibfnamefont {Q.-K.}\ \bibnamefont {Xue}},\
  }\href {http://arxiv.org/abs/1512.01948} {\bibfield  {journal} {\bibinfo
  {journal} {arXiv:1512.01948 [cond-mat]}\ } (\bibinfo {year} {2015})},\
  \bibinfo {note} {arXiv: 1512.01948}\BibitemShut {NoStop}%
\bibitem [{\citenamefont {Zhang}\ \emph {et~al.}(2015)\citenamefont {Zhang},
  \citenamefont {Peng}, \citenamefont {Qian}, \citenamefont {Richard},
  \citenamefont {Shi}, \citenamefont {Ma}, \citenamefont {Fu}, \citenamefont
  {Guo}, \citenamefont {Han}, \citenamefont {Wang}, \citenamefont {Wang},
  \citenamefont {Xue}, \citenamefont {Hu}, \citenamefont {Sun},\ and\
  \citenamefont {Ding}}]{zhang_Ding_2015}%
  \BibitemOpen
  \bibfield  {author} {\bibinfo {author} {\bibfnamefont {P.}~\bibnamefont
  {Zhang}}, \bibinfo {author} {\bibfnamefont {X.-L.}\ \bibnamefont {Peng}},
  \bibinfo {author} {\bibfnamefont {T.}~\bibnamefont {Qian}}, \bibinfo {author}
  {\bibfnamefont {P.}~\bibnamefont {Richard}}, \bibinfo {author} {\bibfnamefont
  {X.}~\bibnamefont {Shi}}, \bibinfo {author} {\bibfnamefont {J.-Z.}\
  \bibnamefont {Ma}}, \bibinfo {author} {\bibfnamefont {B.-B.}\ \bibnamefont
  {Fu}}, \bibinfo {author} {\bibfnamefont {Y.-L.}\ \bibnamefont {Guo}},
  \bibinfo {author} {\bibfnamefont {Z.~Q.}\ \bibnamefont {Han}}, \bibinfo
  {author} {\bibfnamefont {S.~C.}\ \bibnamefont {Wang}}, \bibinfo {author}
  {\bibfnamefont {L.~L.}\ \bibnamefont {Wang}}, \bibinfo {author}
  {\bibfnamefont {Q.-K.}\ \bibnamefont {Xue}}, \bibinfo {author} {\bibfnamefont
  {J.~P.}\ \bibnamefont {Hu}}, \bibinfo {author} {\bibfnamefont {Y.-J.}\
  \bibnamefont {Sun}}, \ and\ \bibinfo {author} {\bibfnamefont
  {H.}~\bibnamefont {Ding}},\ }\href {http://arxiv.org/abs/1512.01949}
  {\bibfield  {journal} {\bibinfo  {journal} {arXiv:1512.01949 [cond-mat]}\ }
  (\bibinfo {year} {2015})},\ \bibinfo {note} {arXiv: 1512.01949}\BibitemShut
  {NoStop}%
\bibitem [{\citenamefont {Miyata}\ \emph {et~al.}(2015)\citenamefont {Miyata},
  \citenamefont {Nakayama}, \citenamefont {Sugawara}, \citenamefont {Sato},\
  and\ \citenamefont {Takahashi}}]{miyata_2015}%
  \BibitemOpen
  \bibfield  {author} {\bibinfo {author} {\bibfnamefont {Y.}~\bibnamefont
  {Miyata}}, \bibinfo {author} {\bibfnamefont {K.}~\bibnamefont {Nakayama}},
  \bibinfo {author} {\bibfnamefont {K.}~\bibnamefont {Sugawara}}, \bibinfo
  {author} {\bibfnamefont {T.}~\bibnamefont {Sato}}, \ and\ \bibinfo {author}
  {\bibfnamefont {T.}~\bibnamefont {Takahashi}},\ }\href {\doibase
  10.1038/nmat4302} {\bibfield  {journal} {\bibinfo  {journal} {Nat Mater}\
  }\textbf {\bibinfo {volume} {14}},\ \bibinfo {pages} {775} (\bibinfo {year}
  {2015})}\BibitemShut {NoStop}%
\bibitem [{\citenamefont {Shiogai}\ \emph {et~al.}(2016)\citenamefont
  {Shiogai}, \citenamefont {Ito}, \citenamefont {Mitsuhashi}, \citenamefont
  {Nojima},\ and\ \citenamefont {Tsukazaki}}]{shiogai_2016}%
  \BibitemOpen
  \bibfield  {author} {\bibinfo {author} {\bibfnamefont {J.}~\bibnamefont
  {Shiogai}}, \bibinfo {author} {\bibfnamefont {Y.}~\bibnamefont {Ito}},
  \bibinfo {author} {\bibfnamefont {T.}~\bibnamefont {Mitsuhashi}}, \bibinfo
  {author} {\bibfnamefont {T.}~\bibnamefont {Nojima}}, \ and\ \bibinfo {author}
  {\bibfnamefont {A.}~\bibnamefont {Tsukazaki}},\ }\href {\doibase
  10.1038/nphys3530} {\bibfield  {journal} {\bibinfo  {journal} {Nat Phys}\
  }\textbf {\bibinfo {volume} {12}},\ \bibinfo {pages} {42} (\bibinfo {year}
  {2016})}\BibitemShut {NoStop}%
\bibitem [{\citenamefont {Wen-Hao}\ \emph {et~al.}(2014)\citenamefont
  {Wen-Hao}, \citenamefont {Yi}, \citenamefont {Jin-Song}, \citenamefont
  {Fang-Sen}, \citenamefont {Ming-Hua}, \citenamefont {Yan-Fei}, \citenamefont
  {Hui-Min}, \citenamefont {Jun-Ping}, \citenamefont {Ying}, \citenamefont
  {Hui-Chao}, \citenamefont {Takeshi}, \citenamefont {Akihiko}, \citenamefont
  {Zhi}, \citenamefont {Hao}, \citenamefont {Chen-Jia}, \citenamefont {Meng},
  \citenamefont {Qing-Yan}, \citenamefont {Ke}, \citenamefont {Shuai-Hua},
  \citenamefont {Xi}, \citenamefont {Jun-Feng}, \citenamefont {Zheng-Cai},
  \citenamefont {Liang}, \citenamefont {Ya-Yu}, \citenamefont {Jian},
  \citenamefont {Li-Li}, \citenamefont {Ming-Wei}, \citenamefont {Qi-Kun},\
  and\ \citenamefont {Xu-Cun}}]{zhang_direct_2014}%
  \BibitemOpen
  \bibfield  {author} {\bibinfo {author} {\bibfnamefont {Z.}~\bibnamefont
  {Wen-Hao}}, \bibinfo {author} {\bibfnamefont {S.}~\bibnamefont {Yi}},
  \bibinfo {author} {\bibfnamefont {Z.}~\bibnamefont {Jin-Song}}, \bibinfo
  {author} {\bibfnamefont {L.}~\bibnamefont {Fang-Sen}}, \bibinfo {author}
  {\bibfnamefont {G.}~\bibnamefont {Ming-Hua}}, \bibinfo {author}
  {\bibfnamefont {Z.}~\bibnamefont {Yan-Fei}}, \bibinfo {author} {\bibfnamefont
  {Z.}~\bibnamefont {Hui-Min}}, \bibinfo {author} {\bibfnamefont
  {P.}~\bibnamefont {Jun-Ping}}, \bibinfo {author} {\bibfnamefont
  {X.}~\bibnamefont {Ying}}, \bibinfo {author} {\bibfnamefont {W.}~\bibnamefont
  {Hui-Chao}}, \bibinfo {author} {\bibfnamefont {F.}~\bibnamefont {Takeshi}},
  \bibinfo {author} {\bibfnamefont {H.}~\bibnamefont {Akihiko}}, \bibinfo
  {author} {\bibfnamefont {L.}~\bibnamefont {Zhi}}, \bibinfo {author}
  {\bibfnamefont {D.}~\bibnamefont {Hao}}, \bibinfo {author} {\bibfnamefont
  {T.}~\bibnamefont {Chen-Jia}}, \bibinfo {author} {\bibfnamefont
  {W.}~\bibnamefont {Meng}}, \bibinfo {author} {\bibfnamefont {W.}~\bibnamefont
  {Qing-Yan}}, \bibinfo {author} {\bibfnamefont {H.}~\bibnamefont {Ke}},
  \bibinfo {author} {\bibfnamefont {J.}~\bibnamefont {Shuai-Hua}}, \bibinfo
  {author} {\bibfnamefont {C.}~\bibnamefont {Xi}}, \bibinfo {author}
  {\bibfnamefont {W.}~\bibnamefont {Jun-Feng}}, \bibinfo {author}
  {\bibfnamefont {X.}~\bibnamefont {Zheng-Cai}}, \bibinfo {author}
  {\bibfnamefont {L.}~\bibnamefont {Liang}}, \bibinfo {author} {\bibfnamefont
  {W.}~\bibnamefont {Ya-Yu}}, \bibinfo {author} {\bibfnamefont
  {W.}~\bibnamefont {Jian}}, \bibinfo {author} {\bibfnamefont {W.}~\bibnamefont
  {Li-Li}}, \bibinfo {author} {\bibfnamefont {C.}~\bibnamefont {Ming-Wei}},
  \bibinfo {author} {\bibfnamefont {X.}~\bibnamefont {Qi-Kun}}, \ and\ \bibinfo
  {author} {\bibfnamefont {M.}~\bibnamefont {Xu-Cun}},\ }\href {\doibase
  10.1088/0256-307X/31/1/017401} {\bibfield  {journal} {\bibinfo  {journal}
  {Chin. Phys. Lett.}\ }\textbf {\bibinfo {volume} {31}},\ \bibinfo {pages}
  {017401} (\bibinfo {year} {2014})}\BibitemShut {NoStop}%
\bibitem [{\citenamefont {Sun}\ \emph {et~al.}(2014)\citenamefont {Sun},
  \citenamefont {Zhang}, \citenamefont {Xing}, \citenamefont {Li},
  \citenamefont {Zhao}, \citenamefont {Xia}, \citenamefont {Wang},
  \citenamefont {Ma}, \citenamefont {Xue},\ and\ \citenamefont
  {Wang}}]{sun_high_2014}%
  \BibitemOpen
  \bibfield  {author} {\bibinfo {author} {\bibfnamefont {Y.}~\bibnamefont
  {Sun}}, \bibinfo {author} {\bibfnamefont {W.}~\bibnamefont {Zhang}}, \bibinfo
  {author} {\bibfnamefont {Y.}~\bibnamefont {Xing}}, \bibinfo {author}
  {\bibfnamefont {F.}~\bibnamefont {Li}}, \bibinfo {author} {\bibfnamefont
  {Y.}~\bibnamefont {Zhao}}, \bibinfo {author} {\bibfnamefont {Z.}~\bibnamefont
  {Xia}}, \bibinfo {author} {\bibfnamefont {L.}~\bibnamefont {Wang}}, \bibinfo
  {author} {\bibfnamefont {X.}~\bibnamefont {Ma}}, \bibinfo {author}
  {\bibfnamefont {Q.-K.}\ \bibnamefont {Xue}}, \ and\ \bibinfo {author}
  {\bibfnamefont {J.}~\bibnamefont {Wang}},\ }\href {\doibase
  10.1038/srep06040} {\bibfield  {journal} {\bibinfo  {journal} {Sci. Rep.}\
  }\textbf {\bibinfo {volume} {4}} (\bibinfo {year} {2014}),\
  10.1038/srep06040}\BibitemShut {NoStop}%
\bibitem [{\citenamefont {Ge}\ \emph {et~al.}(2015)\citenamefont {Ge},
  \citenamefont {Liu}, \citenamefont {Liu}, \citenamefont {Gao}, \citenamefont
  {Qian}, \citenamefont {Xue}, \citenamefont {Liu},\ and\ \citenamefont
  {Jia}}]{ge_2015}%
  \BibitemOpen
  \bibfield  {author} {\bibinfo {author} {\bibfnamefont {J.-F.}\ \bibnamefont
  {Ge}}, \bibinfo {author} {\bibfnamefont {Z.-L.}\ \bibnamefont {Liu}},
  \bibinfo {author} {\bibfnamefont {C.}~\bibnamefont {Liu}}, \bibinfo {author}
  {\bibfnamefont {C.-L.}\ \bibnamefont {Gao}}, \bibinfo {author} {\bibfnamefont
  {D.}~\bibnamefont {Qian}}, \bibinfo {author} {\bibfnamefont {Q.-K.}\
  \bibnamefont {Xue}}, \bibinfo {author} {\bibfnamefont {Y.}~\bibnamefont
  {Liu}}, \ and\ \bibinfo {author} {\bibfnamefont {J.-F.}\ \bibnamefont
  {Jia}},\ }\href {\doibase 10.1038/nmat4153} {\bibfield  {journal} {\bibinfo
  {journal} {Nat Mater}\ }\textbf {\bibinfo {volume} {14}},\ \bibinfo {pages}
  {285} (\bibinfo {year} {2015})}\BibitemShut {NoStop}%
\bibitem [{\citenamefont {Liu}\ \emph {et~al.}(2012{\natexlab{a}})\citenamefont
  {Liu}, \citenamefont {Zhang}, \citenamefont {Mou}, \citenamefont {He},
  \citenamefont {Ou}, \citenamefont {Wang}, \citenamefont {Li}, \citenamefont
  {Wang}, \citenamefont {Zhao}, \citenamefont {He}, \citenamefont {Peng},
  \citenamefont {Liu}, \citenamefont {Chen}, \citenamefont {Yu}, \citenamefont
  {Liu}, \citenamefont {Dong}, \citenamefont {Zhang}, \citenamefont {Chen},
  \citenamefont {Xu}, \citenamefont {Hu}, \citenamefont {Chen}, \citenamefont
  {Ma}, \citenamefont {Xue},\ and\ \citenamefont {Zhou}}]{liu_electronic_2012}%
  \BibitemOpen
  \bibfield  {author} {\bibinfo {author} {\bibfnamefont {D.}~\bibnamefont
  {Liu}}, \bibinfo {author} {\bibfnamefont {W.}~\bibnamefont {Zhang}}, \bibinfo
  {author} {\bibfnamefont {D.}~\bibnamefont {Mou}}, \bibinfo {author}
  {\bibfnamefont {J.}~\bibnamefont {He}}, \bibinfo {author} {\bibfnamefont
  {Y.-B.}\ \bibnamefont {Ou}}, \bibinfo {author} {\bibfnamefont {Q.-Y.}\
  \bibnamefont {Wang}}, \bibinfo {author} {\bibfnamefont {Z.}~\bibnamefont
  {Li}}, \bibinfo {author} {\bibfnamefont {L.}~\bibnamefont {Wang}}, \bibinfo
  {author} {\bibfnamefont {L.}~\bibnamefont {Zhao}}, \bibinfo {author}
  {\bibfnamefont {S.}~\bibnamefont {He}}, \bibinfo {author} {\bibfnamefont
  {Y.}~\bibnamefont {Peng}}, \bibinfo {author} {\bibfnamefont {X.}~\bibnamefont
  {Liu}}, \bibinfo {author} {\bibfnamefont {C.}~\bibnamefont {Chen}}, \bibinfo
  {author} {\bibfnamefont {L.}~\bibnamefont {Yu}}, \bibinfo {author}
  {\bibfnamefont {G.}~\bibnamefont {Liu}}, \bibinfo {author} {\bibfnamefont
  {X.}~\bibnamefont {Dong}}, \bibinfo {author} {\bibfnamefont {J.}~\bibnamefont
  {Zhang}}, \bibinfo {author} {\bibfnamefont {C.}~\bibnamefont {Chen}},
  \bibinfo {author} {\bibfnamefont {Z.}~\bibnamefont {Xu}}, \bibinfo {author}
  {\bibfnamefont {J.}~\bibnamefont {Hu}}, \bibinfo {author} {\bibfnamefont
  {X.}~\bibnamefont {Chen}}, \bibinfo {author} {\bibfnamefont {X.}~\bibnamefont
  {Ma}}, \bibinfo {author} {\bibfnamefont {Q.}~\bibnamefont {Xue}}, \ and\
  \bibinfo {author} {\bibfnamefont {X.~J.}\ \bibnamefont {Zhou}},\ }\href
  {\doibase 10.1038/ncomms1946} {\bibfield  {journal} {\bibinfo  {journal} {Nat
  Commun}\ }\textbf {\bibinfo {volume} {3}},\ \bibinfo {pages} {931} (\bibinfo
  {year} {2012}{\natexlab{a}})}\BibitemShut {NoStop}%
\bibitem [{\citenamefont {He}\ \emph {et~al.}(2013)\citenamefont {He},
  \citenamefont {He}, \citenamefont {Zhang}, \citenamefont {Zhao},
  \citenamefont {Liu}, \citenamefont {Liu}, \citenamefont {Mou}, \citenamefont
  {Ou}, \citenamefont {Wang}, \citenamefont {Li}, \citenamefont {Wang},
  \citenamefont {Peng}, \citenamefont {Liu}, \citenamefont {Chen},
  \citenamefont {Yu}, \citenamefont {Liu}, \citenamefont {Dong}, \citenamefont
  {Zhang}, \citenamefont {Chen}, \citenamefont {Xu}, \citenamefont {Chen},
  \citenamefont {Ma}, \citenamefont {Xue},\ and\ \citenamefont
  {Zhou}}]{he_phase_2013}%
  \BibitemOpen
  \bibfield  {author} {\bibinfo {author} {\bibfnamefont {S.}~\bibnamefont
  {He}}, \bibinfo {author} {\bibfnamefont {J.}~\bibnamefont {He}}, \bibinfo
  {author} {\bibfnamefont {W.}~\bibnamefont {Zhang}}, \bibinfo {author}
  {\bibfnamefont {L.}~\bibnamefont {Zhao}}, \bibinfo {author} {\bibfnamefont
  {D.}~\bibnamefont {Liu}}, \bibinfo {author} {\bibfnamefont {X.}~\bibnamefont
  {Liu}}, \bibinfo {author} {\bibfnamefont {D.}~\bibnamefont {Mou}}, \bibinfo
  {author} {\bibfnamefont {Y.-B.}\ \bibnamefont {Ou}}, \bibinfo {author}
  {\bibfnamefont {Q.-Y.}\ \bibnamefont {Wang}}, \bibinfo {author}
  {\bibfnamefont {Z.}~\bibnamefont {Li}}, \bibinfo {author} {\bibfnamefont
  {L.}~\bibnamefont {Wang}}, \bibinfo {author} {\bibfnamefont {Y.}~\bibnamefont
  {Peng}}, \bibinfo {author} {\bibfnamefont {Y.}~\bibnamefont {Liu}}, \bibinfo
  {author} {\bibfnamefont {C.}~\bibnamefont {Chen}}, \bibinfo {author}
  {\bibfnamefont {L.}~\bibnamefont {Yu}}, \bibinfo {author} {\bibfnamefont
  {G.}~\bibnamefont {Liu}}, \bibinfo {author} {\bibfnamefont {X.}~\bibnamefont
  {Dong}}, \bibinfo {author} {\bibfnamefont {J.}~\bibnamefont {Zhang}},
  \bibinfo {author} {\bibfnamefont {C.}~\bibnamefont {Chen}}, \bibinfo {author}
  {\bibfnamefont {Z.}~\bibnamefont {Xu}}, \bibinfo {author} {\bibfnamefont
  {X.}~\bibnamefont {Chen}}, \bibinfo {author} {\bibfnamefont {X.}~\bibnamefont
  {Ma}}, \bibinfo {author} {\bibfnamefont {Q.}~\bibnamefont {Xue}}, \ and\
  \bibinfo {author} {\bibfnamefont {X.~J.}\ \bibnamefont {Zhou}},\ }\href
  {\doibase 10.1038/nmat3648} {\bibfield  {journal} {\bibinfo  {journal} {Nat
  Mater}\ }\textbf {\bibinfo {volume} {12}},\ \bibinfo {pages} {605} (\bibinfo
  {year} {2013})}\BibitemShut {NoStop}%
\bibitem [{\citenamefont {Tan}\ \emph {et~al.}(2013)\citenamefont {Tan},
  \citenamefont {Zhang}, \citenamefont {Xia}, \citenamefont {Ye}, \citenamefont
  {Chen}, \citenamefont {Xie}, \citenamefont {Peng}, \citenamefont {Xu},
  \citenamefont {Fan}, \citenamefont {Xu}, \citenamefont {Jiang}, \citenamefont
  {Zhang}, \citenamefont {Lai}, \citenamefont {Xiang}, \citenamefont {Hu},
  \citenamefont {Xie},\ and\ \citenamefont {Feng}}]{tan_2013}%
  \BibitemOpen
  \bibfield  {author} {\bibinfo {author} {\bibfnamefont {S.}~\bibnamefont
  {Tan}}, \bibinfo {author} {\bibfnamefont {Y.}~\bibnamefont {Zhang}}, \bibinfo
  {author} {\bibfnamefont {M.}~\bibnamefont {Xia}}, \bibinfo {author}
  {\bibfnamefont {Z.}~\bibnamefont {Ye}}, \bibinfo {author} {\bibfnamefont
  {F.}~\bibnamefont {Chen}}, \bibinfo {author} {\bibfnamefont {X.}~\bibnamefont
  {Xie}}, \bibinfo {author} {\bibfnamefont {R.}~\bibnamefont {Peng}}, \bibinfo
  {author} {\bibfnamefont {D.}~\bibnamefont {Xu}}, \bibinfo {author}
  {\bibfnamefont {Q.}~\bibnamefont {Fan}}, \bibinfo {author} {\bibfnamefont
  {H.}~\bibnamefont {Xu}}, \bibinfo {author} {\bibfnamefont {J.}~\bibnamefont
  {Jiang}}, \bibinfo {author} {\bibfnamefont {T.}~\bibnamefont {Zhang}},
  \bibinfo {author} {\bibfnamefont {X.}~\bibnamefont {Lai}}, \bibinfo {author}
  {\bibfnamefont {T.}~\bibnamefont {Xiang}}, \bibinfo {author} {\bibfnamefont
  {J.}~\bibnamefont {Hu}}, \bibinfo {author} {\bibfnamefont {B.}~\bibnamefont
  {Xie}}, \ and\ \bibinfo {author} {\bibfnamefont {D.}~\bibnamefont {Feng}},\
  }\href {\doibase 10.1038/nmat3654} {\bibfield  {journal} {\bibinfo  {journal}
  {Nat Mater}\ }\textbf {\bibinfo {volume} {12}},\ \bibinfo {pages} {634}
  (\bibinfo {year} {2013})}\BibitemShut {NoStop}%
\bibitem [{\citenamefont {Liu}\ \emph {et~al.}(2014)\citenamefont {Liu},
  \citenamefont {Liu}, \citenamefont {Zhang}, \citenamefont {He}, \citenamefont
  {Zhao}, \citenamefont {He}, \citenamefont {Mou}, \citenamefont {Li},
  \citenamefont {Tang}, \citenamefont {Li}, \citenamefont {Wang}, \citenamefont
  {Peng}, \citenamefont {Liu}, \citenamefont {Chen}, \citenamefont {Yu},
  \citenamefont {Liu}, \citenamefont {Dong}, \citenamefont {Zhang},
  \citenamefont {Chen}, \citenamefont {Xu}, \citenamefont {Chen}, \citenamefont
  {Ma}, \citenamefont {Xue},\ and\ \citenamefont {Zhou}}]{liu_dichotomy_2014}%
  \BibitemOpen
  \bibfield  {author} {\bibinfo {author} {\bibfnamefont {X.}~\bibnamefont
  {Liu}}, \bibinfo {author} {\bibfnamefont {D.}~\bibnamefont {Liu}}, \bibinfo
  {author} {\bibfnamefont {W.}~\bibnamefont {Zhang}}, \bibinfo {author}
  {\bibfnamefont {J.}~\bibnamefont {He}}, \bibinfo {author} {\bibfnamefont
  {L.}~\bibnamefont {Zhao}}, \bibinfo {author} {\bibfnamefont {S.}~\bibnamefont
  {He}}, \bibinfo {author} {\bibfnamefont {D.}~\bibnamefont {Mou}}, \bibinfo
  {author} {\bibfnamefont {F.}~\bibnamefont {Li}}, \bibinfo {author}
  {\bibfnamefont {C.}~\bibnamefont {Tang}}, \bibinfo {author} {\bibfnamefont
  {Z.}~\bibnamefont {Li}}, \bibinfo {author} {\bibfnamefont {L.}~\bibnamefont
  {Wang}}, \bibinfo {author} {\bibfnamefont {Y.}~\bibnamefont {Peng}}, \bibinfo
  {author} {\bibfnamefont {Y.}~\bibnamefont {Liu}}, \bibinfo {author}
  {\bibfnamefont {C.}~\bibnamefont {Chen}}, \bibinfo {author} {\bibfnamefont
  {L.}~\bibnamefont {Yu}}, \bibinfo {author} {\bibfnamefont {G.}~\bibnamefont
  {Liu}}, \bibinfo {author} {\bibfnamefont {X.}~\bibnamefont {Dong}}, \bibinfo
  {author} {\bibfnamefont {J.}~\bibnamefont {Zhang}}, \bibinfo {author}
  {\bibfnamefont {C.}~\bibnamefont {Chen}}, \bibinfo {author} {\bibfnamefont
  {Z.}~\bibnamefont {Xu}}, \bibinfo {author} {\bibfnamefont {X.}~\bibnamefont
  {Chen}}, \bibinfo {author} {\bibfnamefont {X.}~\bibnamefont {Ma}}, \bibinfo
  {author} {\bibfnamefont {Q.}~\bibnamefont {Xue}}, \ and\ \bibinfo {author}
  {\bibfnamefont {X.~J.}\ \bibnamefont {Zhou}},\ }\href {\doibase
  10.1038/ncomms6047} {\bibfield  {journal} {\bibinfo  {journal} {Nat Commun}\
  }\textbf {\bibinfo {volume} {5}} (\bibinfo {year} {2014}),\
  10.1038/ncomms6047}\BibitemShut {NoStop}%
\bibitem [{\citenamefont {Mazin}\ \emph {et~al.}(2008)\citenamefont {Mazin},
  \citenamefont {Singh}, \citenamefont {Johannes},\ and\ \citenamefont
  {Du}}]{mazin_2008}%
  \BibitemOpen
  \bibfield  {author} {\bibinfo {author} {\bibfnamefont {I.}~\bibnamefont
  {Mazin}}, \bibinfo {author} {\bibfnamefont {D.~J.}\ \bibnamefont {Singh}},
  \bibinfo {author} {\bibfnamefont {M.~D.}\ \bibnamefont {Johannes}}, \ and\
  \bibinfo {author} {\bibfnamefont {M.}~\bibnamefont {Du}},\ }\href {\doibase
  10.1103/PhysRevLett.101.057003} {\bibfield  {journal} {\bibinfo  {journal}
  {Phys. Rev. Lett.}\ }\textbf {\bibinfo {volume} {101}},\ \bibinfo {pages}
  {057003} (\bibinfo {year} {2008})}\BibitemShut {NoStop}%
\bibitem [{\citenamefont {Kuroki}\ \emph {et~al.}(2008)\citenamefont {Kuroki},
  \citenamefont {Onari}, \citenamefont {Arita}, \citenamefont {Usui},
  \citenamefont {Tanaka}, \citenamefont {Kontani},\ and\ \citenamefont
  {Aoki}}]{kuroki_2008}%
  \BibitemOpen
  \bibfield  {author} {\bibinfo {author} {\bibfnamefont {K.}~\bibnamefont
  {Kuroki}}, \bibinfo {author} {\bibfnamefont {S.}~\bibnamefont {Onari}},
  \bibinfo {author} {\bibfnamefont {R.}~\bibnamefont {Arita}}, \bibinfo
  {author} {\bibfnamefont {H.}~\bibnamefont {Usui}}, \bibinfo {author}
  {\bibfnamefont {Y.}~\bibnamefont {Tanaka}}, \bibinfo {author} {\bibfnamefont
  {H.}~\bibnamefont {Kontani}}, \ and\ \bibinfo {author} {\bibfnamefont
  {H.}~\bibnamefont {Aoki}},\ }\href {\doibase 10.1103/PhysRevLett.101.087004}
  {\bibfield  {journal} {\bibinfo  {journal} {Phys. Rev. Lett.}\ }\textbf
  {\bibinfo {volume} {101}},\ \bibinfo {pages} {087004} (\bibinfo {year}
  {2008})}\BibitemShut {NoStop}%
\bibitem [{\citenamefont {Peng}\ \emph {et~al.}(2014)\citenamefont {Peng},
  \citenamefont {Xu}, \citenamefont {Tan}, \citenamefont {Cao}, \citenamefont
  {Xia}, \citenamefont {Shen}, \citenamefont {Huang}, \citenamefont {Wen},
  \citenamefont {Song}, \citenamefont {Zhang}, \citenamefont {Xie},
  \citenamefont {Gong},\ and\ \citenamefont {Feng}}]{peng_tuning_2014}%
  \BibitemOpen
  \bibfield  {author} {\bibinfo {author} {\bibfnamefont {R.}~\bibnamefont
  {Peng}}, \bibinfo {author} {\bibfnamefont {H.~C.}\ \bibnamefont {Xu}},
  \bibinfo {author} {\bibfnamefont {S.~Y.}\ \bibnamefont {Tan}}, \bibinfo
  {author} {\bibfnamefont {H.~Y.}\ \bibnamefont {Cao}}, \bibinfo {author}
  {\bibfnamefont {M.}~\bibnamefont {Xia}}, \bibinfo {author} {\bibfnamefont
  {X.~P.}\ \bibnamefont {Shen}}, \bibinfo {author} {\bibfnamefont {Z.~C.}\
  \bibnamefont {Huang}}, \bibinfo {author} {\bibfnamefont {C.~H.~P.}\
  \bibnamefont {Wen}}, \bibinfo {author} {\bibfnamefont {Q.}~\bibnamefont
  {Song}}, \bibinfo {author} {\bibfnamefont {T.}~\bibnamefont {Zhang}},
  \bibinfo {author} {\bibfnamefont {B.~P.}\ \bibnamefont {Xie}}, \bibinfo
  {author} {\bibfnamefont {X.~G.}\ \bibnamefont {Gong}}, \ and\ \bibinfo
  {author} {\bibfnamefont {D.~L.}\ \bibnamefont {Feng}},\ }\href {\doibase
  10.1038/ncomms6044} {\bibfield  {journal} {\bibinfo  {journal} {Nat Commun}\
  }\textbf {\bibinfo {volume} {5}} (\bibinfo {year} {2014}),\
  10.1038/ncomms6044}\BibitemShut {NoStop}%
\bibitem [{\citenamefont {Lee}\ \emph {et~al.}(2014)\citenamefont {Lee},
  \citenamefont {Schmitt}, \citenamefont {Moore}, \citenamefont {Johnston},
  \citenamefont {Cui}, \citenamefont {Li}, \citenamefont {Yi}, \citenamefont
  {Liu}, \citenamefont {Hashimoto}, \citenamefont {Zhang}, \citenamefont {Lu},
  \citenamefont {Devereaux}, \citenamefont {Lee},\ and\ \citenamefont
  {Shen}}]{lee_interfacial_2014}%
  \BibitemOpen
  \bibfield  {author} {\bibinfo {author} {\bibfnamefont {J.~J.}\ \bibnamefont
  {Lee}}, \bibinfo {author} {\bibfnamefont {F.~T.}\ \bibnamefont {Schmitt}},
  \bibinfo {author} {\bibfnamefont {R.~G.}\ \bibnamefont {Moore}}, \bibinfo
  {author} {\bibfnamefont {S.}~\bibnamefont {Johnston}}, \bibinfo {author}
  {\bibfnamefont {Y.-T.}\ \bibnamefont {Cui}}, \bibinfo {author} {\bibfnamefont
  {W.}~\bibnamefont {Li}}, \bibinfo {author} {\bibfnamefont {M.}~\bibnamefont
  {Yi}}, \bibinfo {author} {\bibfnamefont {Z.~K.}\ \bibnamefont {Liu}},
  \bibinfo {author} {\bibfnamefont {M.}~\bibnamefont {Hashimoto}}, \bibinfo
  {author} {\bibfnamefont {Y.}~\bibnamefont {Zhang}}, \bibinfo {author}
  {\bibfnamefont {D.~H.}\ \bibnamefont {Lu}}, \bibinfo {author} {\bibfnamefont
  {T.~P.}\ \bibnamefont {Devereaux}}, \bibinfo {author} {\bibfnamefont {D.-H.}\
  \bibnamefont {Lee}}, \ and\ \bibinfo {author} {\bibfnamefont {Z.-X.}\
  \bibnamefont {Shen}},\ }\href {\doibase 10.1038/nature13894} {\bibfield
  {journal} {\bibinfo  {journal} {Nature}\ }\textbf {\bibinfo {volume} {515}},\
  \bibinfo {pages} {245} (\bibinfo {year} {2014})}\BibitemShut {NoStop}%
\bibitem [{\citenamefont {Zhang}\ \emph {et~al.}(2014)\citenamefont {Zhang},
  \citenamefont {Li}, \citenamefont {Li}, \citenamefont {Zhang}, \citenamefont
  {Peng}, \citenamefont {Tang}, \citenamefont {Wang}, \citenamefont {He},
  \citenamefont {Chen}, \citenamefont {Wang}, \citenamefont {Ma},\ and\
  \citenamefont {Xue}}]{zhang_Xue_2014}%
  \BibitemOpen
  \bibfield  {author} {\bibinfo {author} {\bibfnamefont {W.}~\bibnamefont
  {Zhang}}, \bibinfo {author} {\bibfnamefont {Z.}~\bibnamefont {Li}}, \bibinfo
  {author} {\bibfnamefont {F.}~\bibnamefont {Li}}, \bibinfo {author}
  {\bibfnamefont {H.}~\bibnamefont {Zhang}}, \bibinfo {author} {\bibfnamefont
  {J.}~\bibnamefont {Peng}}, \bibinfo {author} {\bibfnamefont {C.}~\bibnamefont
  {Tang}}, \bibinfo {author} {\bibfnamefont {Q.}~\bibnamefont {Wang}}, \bibinfo
  {author} {\bibfnamefont {K.}~\bibnamefont {He}}, \bibinfo {author}
  {\bibfnamefont {X.}~\bibnamefont {Chen}}, \bibinfo {author} {\bibfnamefont
  {L.}~\bibnamefont {Wang}}, \bibinfo {author} {\bibfnamefont {X.}~\bibnamefont
  {Ma}}, \ and\ \bibinfo {author} {\bibfnamefont {Q.-K.}\ \bibnamefont {Xue}},\
  }\href {\doibase 10.1103/PhysRevB.89.060506} {\bibfield  {journal} {\bibinfo
  {journal} {Phys. Rev. B}\ }\textbf {\bibinfo {volume} {89}},\ \bibinfo
  {pages} {060506} (\bibinfo {year} {2014})}\BibitemShut {NoStop}%
\bibitem [{\citenamefont {Bazhirov}\ and\ \citenamefont
  {Cohen}(2013)}]{bazhirov_2013}%
  \BibitemOpen
  \bibfield  {author} {\bibinfo {author} {\bibfnamefont {T.}~\bibnamefont
  {Bazhirov}}\ and\ \bibinfo {author} {\bibfnamefont {M.~L.}\ \bibnamefont
  {Cohen}},\ }\href {\doibase 10.1088/0953-8984/25/10/105506} {\bibfield
  {journal} {\bibinfo  {journal} {J. Phys.: Condens. Matter}\ }\textbf
  {\bibinfo {volume} {25}},\ \bibinfo {pages} {105506} (\bibinfo {year}
  {2013})}\BibitemShut {NoStop}%
\bibitem [{\citenamefont {Liu}\ \emph {et~al.}(2012{\natexlab{b}})\citenamefont
  {Liu}, \citenamefont {Lu},\ and\ \citenamefont {Xiang}}]{liu_atomic_2012}%
  \BibitemOpen
  \bibfield  {author} {\bibinfo {author} {\bibfnamefont {K.}~\bibnamefont
  {Liu}}, \bibinfo {author} {\bibfnamefont {Z.-Y.}\ \bibnamefont {Lu}}, \ and\
  \bibinfo {author} {\bibfnamefont {T.}~\bibnamefont {Xiang}},\ }\href
  {\doibase 10.1103/PhysRevB.85.235123} {\bibfield  {journal} {\bibinfo
  {journal} {Phys. Rev. B}\ }\textbf {\bibinfo {volume} {85}},\ \bibinfo
  {pages} {235123} (\bibinfo {year} {2012}{\natexlab{b}})}\BibitemShut
  {NoStop}%
\bibitem [{\citenamefont {Bang}\ \emph {et~al.}(2013)\citenamefont {Bang},
  \citenamefont {Li}, \citenamefont {Sun}, \citenamefont {Samanta},
  \citenamefont {Zhang}, \citenamefont {Zhang}, \citenamefont {Wang},
  \citenamefont {Chen}, \citenamefont {Ma}, \citenamefont {Xue},\ and\
  \citenamefont {Zhang}}]{bang_atomic_2013}%
  \BibitemOpen
  \bibfield  {author} {\bibinfo {author} {\bibfnamefont {J.}~\bibnamefont
  {Bang}}, \bibinfo {author} {\bibfnamefont {Z.}~\bibnamefont {Li}}, \bibinfo
  {author} {\bibfnamefont {Y.~Y.}\ \bibnamefont {Sun}}, \bibinfo {author}
  {\bibfnamefont {A.}~\bibnamefont {Samanta}}, \bibinfo {author} {\bibfnamefont
  {Y.~Y.}\ \bibnamefont {Zhang}}, \bibinfo {author} {\bibfnamefont
  {W.}~\bibnamefont {Zhang}}, \bibinfo {author} {\bibfnamefont
  {L.}~\bibnamefont {Wang}}, \bibinfo {author} {\bibfnamefont {X.}~\bibnamefont
  {Chen}}, \bibinfo {author} {\bibfnamefont {X.}~\bibnamefont {Ma}}, \bibinfo
  {author} {\bibfnamefont {Q.-K.}\ \bibnamefont {Xue}}, \ and\ \bibinfo
  {author} {\bibfnamefont {S.~B.}\ \bibnamefont {Zhang}},\ }\href {\doibase
  10.1103/PhysRevB.87.220503} {\bibfield  {journal} {\bibinfo  {journal} {Phys.
  Rev. B}\ }\textbf {\bibinfo {volume} {87}},\ \bibinfo {pages} {220503}
  (\bibinfo {year} {2013})}\BibitemShut {NoStop}%
\bibitem [{\citenamefont {Cao}\ \emph {et~al.}(2014)\citenamefont {Cao},
  \citenamefont {Tan}, \citenamefont {Xiang}, \citenamefont {Feng},\ and\
  \citenamefont {Gong}}]{cao_interfacial_2014}%
  \BibitemOpen
  \bibfield  {author} {\bibinfo {author} {\bibfnamefont {H.-Y.}\ \bibnamefont
  {Cao}}, \bibinfo {author} {\bibfnamefont {S.}~\bibnamefont {Tan}}, \bibinfo
  {author} {\bibfnamefont {H.}~\bibnamefont {Xiang}}, \bibinfo {author}
  {\bibfnamefont {D.~L.}\ \bibnamefont {Feng}}, \ and\ \bibinfo {author}
  {\bibfnamefont {X.-G.}\ \bibnamefont {Gong}},\ }\href {\doibase
  10.1103/PhysRevB.89.014501} {\bibfield  {journal} {\bibinfo  {journal} {Phys.
  Rev. B}\ }\textbf {\bibinfo {volume} {89}},\ \bibinfo {pages} {014501}
  (\bibinfo {year} {2014})}\BibitemShut {NoStop}%
\bibitem [{\citenamefont {Xie}\ \emph {et~al.}(2015)\citenamefont {Xie},
  \citenamefont {Cao}, \citenamefont {Zhou}, \citenamefont {Chen},
  \citenamefont {Xiang},\ and\ \citenamefont {Gong}}]{xie_oxygen_2015}%
  \BibitemOpen
  \bibfield  {author} {\bibinfo {author} {\bibfnamefont {Y.}~\bibnamefont
  {Xie}}, \bibinfo {author} {\bibfnamefont {H.-Y.}\ \bibnamefont {Cao}},
  \bibinfo {author} {\bibfnamefont {Y.}~\bibnamefont {Zhou}}, \bibinfo {author}
  {\bibfnamefont {S.}~\bibnamefont {Chen}}, \bibinfo {author} {\bibfnamefont
  {H.}~\bibnamefont {Xiang}}, \ and\ \bibinfo {author} {\bibfnamefont {X.-G.}\
  \bibnamefont {Gong}},\ }\href {http://dx.doi.org/10.1038/srep10011}
  {\bibfield  {journal} {\bibinfo  {journal} {Sci. Rep.}\ }\textbf {\bibinfo
  {volume} {5}},\ \bibinfo {pages} {10011} (\bibinfo {year}
  {2015})}\BibitemShut {NoStop}%
\bibitem [{\citenamefont {Shanavas}\ and\ \citenamefont
  {Singh}(2015)}]{shanavas_2015}%
  \BibitemOpen
  \bibfield  {author} {\bibinfo {author} {\bibfnamefont {K.~V.}\ \bibnamefont
  {Shanavas}}\ and\ \bibinfo {author} {\bibfnamefont {D.~J.}\ \bibnamefont
  {Singh}},\ }\href {\doibase 10.1103/PhysRevB.92.035144} {\bibfield  {journal}
  {\bibinfo  {journal} {Phys. Rev. B}\ }\textbf {\bibinfo {volume} {92}},\
  \bibinfo {pages} {035144} (\bibinfo {year} {2015})}\BibitemShut {NoStop}%
\bibitem [{\citenamefont {Zheng}\ \emph {et~al.}(2015)\citenamefont {Zheng},
  \citenamefont {Wang}, \citenamefont {Xue},\ and\ \citenamefont
  {Zhang}}]{zheng_band_2015}%
  \BibitemOpen
  \bibfield  {author} {\bibinfo {author} {\bibfnamefont {F.}~\bibnamefont
  {Zheng}}, \bibinfo {author} {\bibfnamefont {L.-L.}\ \bibnamefont {Wang}},
  \bibinfo {author} {\bibfnamefont {Q.-K.}\ \bibnamefont {Xue}}, \ and\
  \bibinfo {author} {\bibfnamefont {P.}~\bibnamefont {Zhang}},\ }\href
  {http://arxiv.org/abs/1508.06498} {\bibfield  {journal} {\bibinfo  {journal}
  {arXiv:1508.06498 [cond-mat]}\ } (\bibinfo {year} {2015})},\ \bibinfo {note}
  {arXiv: 1508.06498}\BibitemShut {NoStop}%
\bibitem [{\citenamefont {Berlijn}\ \emph {et~al.}(2014)\citenamefont
  {Berlijn}, \citenamefont {Cheng}, \citenamefont {Hirschfeld},\ and\
  \citenamefont {Ku}}]{berlijn_2014}%
  \BibitemOpen
  \bibfield  {author} {\bibinfo {author} {\bibfnamefont {T.}~\bibnamefont
  {Berlijn}}, \bibinfo {author} {\bibfnamefont {H.-P.}\ \bibnamefont {Cheng}},
  \bibinfo {author} {\bibfnamefont {P.~J.}\ \bibnamefont {Hirschfeld}}, \ and\
  \bibinfo {author} {\bibfnamefont {W.}~\bibnamefont {Ku}},\ }\href {\doibase
  10.1103/PhysRevB.89.020501} {\bibfield  {journal} {\bibinfo  {journal} {Phys.
  Rev. B}\ }\textbf {\bibinfo {volume} {89}},\ \bibinfo {pages} {020501}
  (\bibinfo {year} {2014})}\BibitemShut {NoStop}%
\bibitem [{\citenamefont {Zheng}\ \emph {et~al.}(2013)\citenamefont {Zheng},
  \citenamefont {Wang}, \citenamefont {Kang},\ and\ \citenamefont
  {Zhang}}]{zheng_2013}%
  \BibitemOpen
  \bibfield  {author} {\bibinfo {author} {\bibfnamefont {F.}~\bibnamefont
  {Zheng}}, \bibinfo {author} {\bibfnamefont {Z.}~\bibnamefont {Wang}},
  \bibinfo {author} {\bibfnamefont {W.}~\bibnamefont {Kang}}, \ and\ \bibinfo
  {author} {\bibfnamefont {P.}~\bibnamefont {Zhang}},\ }\href
  {http://dx.doi.org/10.1038/srep02213} {\bibfield  {journal} {\bibinfo
  {journal} {Scientific Reports}\ }\textbf {\bibinfo {volume} {3}},\ \bibinfo
  {pages} {2213} (\bibinfo {year} {2013})}\BibitemShut {NoStop}%
\bibitem [{\citenamefont {Liu}\ \emph {et~al.}(2015)\citenamefont {Liu},
  \citenamefont {Zhang},\ and\ \citenamefont {Lu}}]{liu_2015}%
  \BibitemOpen
  \bibfield  {author} {\bibinfo {author} {\bibfnamefont {K.}~\bibnamefont
  {Liu}}, \bibinfo {author} {\bibfnamefont {B.-J.}\ \bibnamefont {Zhang}}, \
  and\ \bibinfo {author} {\bibfnamefont {Z.-Y.}\ \bibnamefont {Lu}},\ }\href
  {\doibase 10.1103/PhysRevB.91.045107} {\bibfield  {journal} {\bibinfo
  {journal} {Phys. Rev. B}\ }\textbf {\bibinfo {volume} {91}},\ \bibinfo
  {pages} {045107} (\bibinfo {year} {2015})}\BibitemShut {NoStop}%
\bibitem [{\citenamefont {Johnson}\ \emph {et~al.}(2015)\citenamefont
  {Johnson}, \citenamefont {Yang}, \citenamefont {Rameau}, \citenamefont {Gu},
  \citenamefont {Pan}, \citenamefont {Valla}, \citenamefont {Weinert},\ and\
  \citenamefont {Fedorov}}]{soc-2015}%
  \BibitemOpen
  \bibfield  {author} {\bibinfo {author} {\bibfnamefont {P.~D.}\ \bibnamefont
  {Johnson}}, \bibinfo {author} {\bibfnamefont {H.-B.}\ \bibnamefont {Yang}},
  \bibinfo {author} {\bibfnamefont {J.~D.}\ \bibnamefont {Rameau}}, \bibinfo
  {author} {\bibfnamefont {G.~D.}\ \bibnamefont {Gu}}, \bibinfo {author}
  {\bibfnamefont {Z.-H.}\ \bibnamefont {Pan}}, \bibinfo {author} {\bibfnamefont
  {T.}~\bibnamefont {Valla}}, \bibinfo {author} {\bibfnamefont
  {M.}~\bibnamefont {Weinert}}, \ and\ \bibinfo {author} {\bibfnamefont
  {A.~V.}\ \bibnamefont {Fedorov}},\ }\href {\doibase
  10.1103/PhysRevLett.114.167001} {\bibfield  {journal} {\bibinfo  {journal}
  {Phys. Rev. Lett.}\ }\textbf {\bibinfo {volume} {114}},\ \bibinfo {pages}
  {167001} (\bibinfo {year} {2015})}\BibitemShut {NoStop}%
\bibitem [{\citenamefont {Kresse}\ and\ \citenamefont
  {Furthm\"uller}(1996{\natexlab{a}})}]{kresse_efficiency_1996}%
  \BibitemOpen
  \bibfield  {author} {\bibinfo {author} {\bibfnamefont {G.}~\bibnamefont
  {Kresse}}\ and\ \bibinfo {author} {\bibfnamefont {J.}~\bibnamefont
  {Furthm\"uller}},\ }\href {\doibase 10.1016/0927-0256(96)00008-0} {\bibfield
  {journal} {\bibinfo  {journal} {Comput. Mater. Sci.}\ }\textbf {\bibinfo
  {volume} {6}},\ \bibinfo {pages} {15} (\bibinfo {year}
  {1996}{\natexlab{a}})}\BibitemShut {NoStop}%
\bibitem [{\citenamefont {Kresse}\ and\ \citenamefont
  {Furthm\"uller}(1996{\natexlab{b}})}]{kresse_efficient_1996}%
  \BibitemOpen
  \bibfield  {author} {\bibinfo {author} {\bibfnamefont {G.}~\bibnamefont
  {Kresse}}\ and\ \bibinfo {author} {\bibfnamefont {J.}~\bibnamefont
  {Furthm\"uller}},\ }\href {\doibase 10.1103/PhysRevB.54.11169} {\bibfield
  {journal} {\bibinfo  {journal} {Phys. Rev. B}\ }\textbf {\bibinfo {volume}
  {54}},\ \bibinfo {pages} {11169} (\bibinfo {year}
  {1996}{\natexlab{b}})}\BibitemShut {NoStop}%
\bibitem [{\citenamefont {Perdew}\ \emph {et~al.}(1996)\citenamefont {Perdew},
  \citenamefont {Burke},\ and\ \citenamefont {Ernzerhof}}]{perdew_1996}%
  \BibitemOpen
  \bibfield  {author} {\bibinfo {author} {\bibfnamefont {J.~P.}\ \bibnamefont
  {Perdew}}, \bibinfo {author} {\bibfnamefont {K.}~\bibnamefont {Burke}}, \
  and\ \bibinfo {author} {\bibfnamefont {M.}~\bibnamefont {Ernzerhof}},\ }\href
  {\doibase 10.1103/PhysRevLett.77.3865} {\bibfield  {journal} {\bibinfo
  {journal} {Phys. Rev. Lett.}\ }\textbf {\bibinfo {volume} {77}},\ \bibinfo
  {pages} {3865} (\bibinfo {year} {1996})}\BibitemShut {NoStop}%
\bibitem [{\citenamefont {Bl\"{o}chl}(1994)}]{bloechl_projector_1994}%
  \BibitemOpen
  \bibfield  {author} {\bibinfo {author} {\bibfnamefont {P.~E.}\ \bibnamefont
  {Bl\"{o}chl}},\ }\href {\doibase 10.1103/PhysRevB.50.17953} {\bibfield
  {journal} {\bibinfo  {journal} {Phys. Rev. B}\ }\textbf {\bibinfo {volume}
  {50}},\ \bibinfo {pages} {17953} (\bibinfo {year} {1994})}\BibitemShut
  {NoStop}%
\bibitem [{\citenamefont {Kresse}\ and\ \citenamefont
  {Joubert}(1999)}]{kresse_ultrasoft_1999}%
  \BibitemOpen
  \bibfield  {author} {\bibinfo {author} {\bibfnamefont {G.}~\bibnamefont
  {Kresse}}\ and\ \bibinfo {author} {\bibfnamefont {D.}~\bibnamefont
  {Joubert}},\ }\href {\doibase 10.1103/PhysRevB.59.1758} {\bibfield  {journal}
  {\bibinfo  {journal} {Phys. Rev. B}\ }\textbf {\bibinfo {volume} {59}},\
  \bibinfo {pages} {1758} (\bibinfo {year} {1999})}\BibitemShut {NoStop}%
\bibitem [{\citenamefont {Klime\ifmmode\check{s}\else\v{s}\fi{}}\ \emph
  {et~al.}(2011)\citenamefont {Klime\ifmmode\check{s}\else\v{s}\fi{}},
  \citenamefont {Bowler},\ and\ \citenamefont
  {Michaelides}}]{PhysRevB.83.195131}%
  \BibitemOpen
  \bibfield  {author} {\bibinfo {author} {\bibfnamefont {J.}~\bibnamefont
  {Klime\ifmmode\check{s}\else\v{s}\fi{}}}, \bibinfo {author} {\bibfnamefont
  {D.~R.}\ \bibnamefont {Bowler}}, \ and\ \bibinfo {author} {\bibfnamefont
  {A.}~\bibnamefont {Michaelides}},\ }\href {\doibase
  10.1103/PhysRevB.83.195131} {\bibfield  {journal} {\bibinfo  {journal} {Phys.
  Rev. B}\ }\textbf {\bibinfo {volume} {83}},\ \bibinfo {pages} {195131}
  (\bibinfo {year} {2011})},\ \bibinfo {note} {the optPBE-vdW method is used in
  the present calculations.}\BibitemShut {Stop}%
\bibitem [{\citenamefont {Qi}\ \emph {et~al.}(2010)\citenamefont {Qi},
  \citenamefont {Rhim}, \citenamefont {Sun}, \citenamefont {Weinert},\ and\
  \citenamefont {Li}}]{bufferlayer}%
  \BibitemOpen
  \bibfield  {author} {\bibinfo {author} {\bibfnamefont {Y.}~\bibnamefont
  {Qi}}, \bibinfo {author} {\bibfnamefont {S.~H.}\ \bibnamefont {Rhim}},
  \bibinfo {author} {\bibfnamefont {G.~F.}\ \bibnamefont {Sun}}, \bibinfo
  {author} {\bibfnamefont {M.}~\bibnamefont {Weinert}}, \ and\ \bibinfo
  {author} {\bibfnamefont {L.}~\bibnamefont {Li}},\ }\href {\doibase
  10.1103/PhysRevLett.105.085502} {\bibfield  {journal} {\bibinfo  {journal}
  {Phys. Rev. Lett.}\ }\textbf {\bibinfo {volume} {105}},\ \bibinfo {pages}
  {085502} (\bibinfo {year} {2010})}\BibitemShut {NoStop}%
\bibitem [{\citenamefont {Goodwin}(1939)}]{Goodwin_39}%
  \BibitemOpen
  \bibfield  {author} {\bibinfo {author} {\bibfnamefont {E.~T.}\ \bibnamefont
  {Goodwin}},\ }\href {https://doi.org/10.1017/S0305004100020910} {\bibfield
  {journal} {\bibinfo  {journal} {Proc. Cambridge Phil. Soc.}\ }\textbf
  {\bibinfo {volume} {35}},\ \bibinfo {pages} {205, 221, 232} (\bibinfo {year}
  {1939})}\BibitemShut {NoStop}%
\bibitem [{\citenamefont {Davenport}\ \emph {et~al.}(1988)\citenamefont
  {Davenport}, \citenamefont {Watson},\ and\ \citenamefont
  {Weinert}}]{kproj_88}%
  \BibitemOpen
  \bibfield  {author} {\bibinfo {author} {\bibfnamefont {J.~W.}\ \bibnamefont
  {Davenport}}, \bibinfo {author} {\bibfnamefont {R.~E.}\ \bibnamefont
  {Watson}}, \ and\ \bibinfo {author} {\bibfnamefont {M.}~\bibnamefont
  {Weinert}},\ }\href {http://link.aps.org/doi/10.1103/PhysRevB.37.9985}
  {\bibfield  {journal} {\bibinfo  {journal} {Phys. Rev. B}\ }\textbf {\bibinfo
  {volume} {37}},\ \bibinfo {pages} {9985} (\bibinfo {year}
  {1988})}\BibitemShut {NoStop}%
\bibitem [{\citenamefont {Chen}\ and\ \citenamefont
  {Weinert}(2014)}]{chen_revealing_2014}%
  \BibitemOpen
  \bibfield  {author} {\bibinfo {author} {\bibfnamefont {M.~X.}\ \bibnamefont
  {Chen}}\ and\ \bibinfo {author} {\bibfnamefont {M.}~\bibnamefont {Weinert}},\
  }\href {\doibase 10.1021/nl502107v} {\bibfield  {journal} {\bibinfo
  {journal} {Nano. Lett.}\ }\textbf {\bibinfo {volume} {14}},\ \bibinfo {pages}
  {5189} (\bibinfo {year} {2014})}\BibitemShut {NoStop}%
\bibitem [{\citenamefont {Borisenko}\ \emph {et~al.}(2016)\citenamefont
  {Borisenko}, \citenamefont {Evtushinsky}, \citenamefont {Liu}, \citenamefont
  {Morozov}, \citenamefont {Kappenberger}, \citenamefont {Wurmehl},
  \citenamefont {Buchner}, \citenamefont {Yaresko}, \citenamefont {Kim},
  \citenamefont {Hoesch}, \citenamefont {Wolf},\ and\ \citenamefont
  {Zhigadlo}}]{borisenko_direct_2016}%
  \BibitemOpen
  \bibfield  {author} {\bibinfo {author} {\bibfnamefont {S.~V.}\ \bibnamefont
  {Borisenko}}, \bibinfo {author} {\bibfnamefont {D.~V.}\ \bibnamefont
  {Evtushinsky}}, \bibinfo {author} {\bibfnamefont {Z.-H.}\ \bibnamefont
  {Liu}}, \bibinfo {author} {\bibfnamefont {I.}~\bibnamefont {Morozov}},
  \bibinfo {author} {\bibfnamefont {R.}~\bibnamefont {Kappenberger}}, \bibinfo
  {author} {\bibfnamefont {S.}~\bibnamefont {Wurmehl}}, \bibinfo {author}
  {\bibfnamefont {B.}~\bibnamefont {Buchner}}, \bibinfo {author} {\bibfnamefont
  {A.~N.}\ \bibnamefont {Yaresko}}, \bibinfo {author} {\bibfnamefont {T.~K.}\
  \bibnamefont {Kim}}, \bibinfo {author} {\bibfnamefont {M.}~\bibnamefont
  {Hoesch}}, \bibinfo {author} {\bibfnamefont {T.}~\bibnamefont {Wolf}}, \ and\
  \bibinfo {author} {\bibfnamefont {N.~D.}\ \bibnamefont {Zhigadlo}},\ }\href
  {http://dx.doi.org/10.1038/nphys3594} {\bibfield  {journal} {\bibinfo
  {journal} {Nat Phys}\ }\textbf {\bibinfo {volume} {12}},\ \bibinfo {pages}
  {311} (\bibinfo {year} {2016})}\BibitemShut {NoStop}%
\end{thebibliography}%
\bibliographystyle{apsrev4-1}
\end{document}